\documentclass[english,aps,pre,reprint,superscriptaddress,showpacs]{revtex4-1}

\usepackage[T1]{fontenc}
\usepackage[latin9]{inputenc}
\usepackage{amsmath}
\usepackage{amssymb}
\usepackage{bbold}
\usepackage{graphicx}
\usepackage{booktabs}
\usepackage{esint}
\usepackage{babel}
\usepackage{color}

\DeclareMathOperator{\const}{const}
\DeclareMathOperator{\e}{e}
\DeclareMathOperator{\eig}{eig}

\begin{document}

\title{Strong and Weak Chaos in Networks of Semiconductor Lasers\\
  with Time-delayed Couplings}

\author{Sven Heiligenthal}

\email{sven.heiligenthal@physik.uni-wuerzburg.de}

\affiliation{Institute of Theoretical Physics, University of
  W\"urzburg, 97074 W\"urzburg, Germany}

\author{Thomas J\"ungling}

\affiliation{Institute of Theoretical Physics, University of
  W\"urzburg, 97074 W\"urzburg, Germany}

\author{Otti D'Huys}

\affiliation{Institute of Theoretical Physics, University of
  W\"urzburg, 97074 W\"urzburg, Germany}

\author{Diana A. Arroyo-Almanza}

\altaffiliation{Also with Centro de Investigaciones en \'Optica,
  Le\'on 37150, Guanajuato, M\'exico.}

\affiliation{Instituto de F\'isica Interdisciplinar y Sistemas
  Complejos, IFISC (UIB-CSIC), Campus Universitat de les Illes
  Balears, 07122 Palma de Mallorca, Spain}

\author{Miguel C. Soriano}

\affiliation{Instituto de F\'isica Interdisciplinar y Sistemas
  Complejos, IFISC (UIB-CSIC), Campus Universitat de les Illes
  Balears, 07122 Palma de Mallorca, Spain}

\author{Ingo Fischer}

\affiliation{Instituto de F\'isica Interdisciplinar y Sistemas
  Complejos, IFISC (UIB-CSIC), Campus Universitat de les Illes
  Balears, 07122 Palma de Mallorca, Spain}

\author{Ido Kanter}

\affiliation{Department of Physics, Bar-Ilan University, 52900
  Ramat-Gan, Israel}

\author{Wolfgang Kinzel}

\affiliation{Institute of Theoretical Physics, University of
  W\"urzburg, 97074 W\"urzburg, Germany}

\date{\today}

\begin{abstract}
  Nonlinear networks with time-delayed couplings may show strong and
  weak chaos, depending on the scaling of their Lyapunov exponent with
  the delay time. We study strong and weak chaos for semiconductor
  lasers, either with time-delayed self-feedback or for small
  networks. We examine the dependence on the pump current and consider
  the question whether strong and weak chaos can be identified from
  the shape of the intensity trace, the auto-correlations and the
  external cavity modes. The concept of the sub-Lyapunov exponent
  $\lambda_0$ is generalized to the case of two time-scale separated
  delays in the system. We give the first experimental evidence of
  strong and weak chaos in a network of lasers which supports the
  sequence `weak to strong to weak chaos' upon monotonically
  increasing the coupling strength. Finally, we discuss strong and
  weak chaos for networks with several distinct sub-Lyapunov exponents
  and comment on the dependence of the sub-Lyapunov exponent on the
  number of a laser's inputs in a network.
\end{abstract}


\maketitle

\section{Introduction}

The cooperative behavior of a system of interacting units is of
fundamental interest in nonlinear dynamics. Such complex networks have
a wide range of interdisciplinary applications ranging from neural
networks to coupled lasers
\cite{Albert2002,Boccaletti2006,Arenas2008,Song2010,Vardi2012,*Kanter2011}.
Typically, these units interact by transmitting information about
their state to their partners, and in many systems the transmission
time is larger than the time scales of the individual units. For this
reason, networks with time-delayed couplings are a focus of active
research \cite{Lakshmanan2010,Just2010}.

Time-delayed feedback can produce dynamical instabilities which may
lead to deterministic chaos
\cite{Farmer1982,Lepri1994,Giacomelli1996}.  Even a scalar nonlinear
differential equation with time-delayed feedback has an
infinite-dimensional phase space which favors chaotic solutions.  As
an example, a single semiconductor laser often exhibits chaotic
emission dynamics when its laser beam is reflected back into its
cavity by an external mirror. Networks of nonlinear units may,
similarly, become chaotic due to time-delayed coupling of the nodes
\cite{Erneux2009}.

For networks of identical units, chaos synchronization is possible.
Even if the delay times are very long, the units may synchronize onto
a common chaotic trajectory without time shift (zero-lag
synchronization) \cite{Klein2006,Fischer2006,Flunkert2010}. Other
kinds of synchronization are possible as well, like phase, achronal,
anticipated and generalized synchronization. Chaos synchronization is
being discussed in the context of secure communication
\cite{Argyris2005,Kanter2008}.

Recently, two different kinds of chaos have been identified for
chaotic networks of time-continuous systems with time-delayed
couplings: strong and weak chaos \cite{Heiligenthal2011}. In the limit
of large delay times, the maximal Lyapunov exponent (LE) of the
network saturates at a nonzero value for strong chaos, whereas it
scales with the inverse delay time for weak chaos. A similar
phenomenon has been reported in \cite{Lepri1994} for time-discrete
maps with delay. Only networks exhibiting weak chaos can synchronize
to a common chaotic trajectory.

In this paper we extend these investigations on strong and weak chaos
focusing on the dynamics of semiconductor lasers. In Sec.~II we define
the model and the linearized equations which determine the Lyapunov
exponents. In Sec.~III a single unit with time-delayed feedback is
considered. An artificial sub-LE is defined which determines, whether
the unit is in the strong or weak chaos phase. The scaling argument of
\cite{Heiligenthal2011} is extended to derive the behavior of the
maximal LE in the limit of large delay time. This system is
investigated in Sec.~IV for semiconductor lasers. Numerical
simulations of the Lang-Kobayashi equations yield the transition from
weak to strong chaos and back to weak chaos upon monotonically
increasing the coupling strength. The scaling just at the transition
is derived. Auto-correlations, spatial representations of the chaotic
intensity and external cavity modes are calculated to investigate
whether one can deduce the type of chaos from a single trajectory. In
Sec.~V networks of semiconductor lasers are considered. The stability
of the synchronization manifold is related to the eigenvalue gap of
the coupling matrix and the master stability function. The first
experiment on semiconductor lasers which supports the sequence `weak
to strong to weak chaos' with increasing coupling strength is
demonstrated. Finally, networks with several distinct sub-LEs and
certain network patterns are investigated.

\section{The Lang-Kobayashi equations and their Lyapunov exponents}

We study the dynamics of coupled semiconductor lasers by means of
numerical simulations, complemented by experiments. To this end we use
the Lang-Kobayashi (LK) model, which compresses the complex physical
processes of a laser in rate equations for only three classical
degrees of freedom, namely the complex electric field $\mathcal{E}(t)$
and the normalized deviation of the charge carrier density $n(t)$ with
respect to the density at the lasing threshold. The rate equations for
a single laser without self-feedback read
\begin{equation}
  \begin{split}
    \dot{\mathcal{E}}(t)&=\frac{1+\mathrm{i}\,\alpha}{2}\,G_{\mathrm{N}}\,n(t)\,\mathcal{E}(t)\\
    \dot{n}(t)&=(p-1)\,J_{\mathrm{th}}-\gamma\,n(t)-\left[\Gamma+G_{\mathrm{N}}\,
      n(t)\right]|\mathcal{E}(t)|^2
  \end{split}
  \label{LKmodel}
\end{equation}
A table of the parameters involved can be found in the appendix. Note
that the phase of the electric field $\phi_e(t)=\arg\mathcal{E}(t)$ in
the model is defined to describe only the deviation from
$\omega_0\,t$, which contains the high oscillation frequency of the
actual laser mode. Moreover, we have omitted intensity dependent
nonlinear gain reduction, which would be taken into account via an
additional factor $(1+\varepsilon\,|\mathcal{E}|^2)^{-1}$ with
$\varepsilon \sim 10^{-7}$ for $G_{\mathrm{N}}$. All results presented
in this paper, which are obtained with the simplified
Eq.~(\ref{LKmodel}), coincide qualitatively with the results of the
complete rate equations. The LEs are calculated by linearizing
Eq.~(\ref{LKmodel}). The linearization reads
\begin{equation}
  \begin{split}
    \dot{\delta \mathcal{E}}(t)=&\frac{1+\mathrm{i}\,\alpha}{2}\,G_{\mathrm{N}}\left[n(t)\,\delta \mathcal{E}(t)+\mathcal{E}(t)\,\delta n(t)\right]\\
    \dot{\delta
      n}(t)=&-\left(\gamma+G_\mathrm{N}\,|\mathcal{E}(t)|^2\right)\delta
    n(t)\\&-2\left[\Gamma+G_\mathrm{N}\,
      n(t)\right]\Re\left\{\mathcal{E}^*(t)\,\delta
      \mathcal{E}(t)\right\}.
  \end{split}
  \label{LKlinearization}
\end{equation}
We describe the system's state by the state vector
$\mathbf{x}=(x_1,x_2,x_3)^\top$ with $x_1\equiv\Re\{\mathcal{E}\}$,
$x_2\equiv\Im\{\mathcal{E}\}$, and $x_3\equiv n$. Consequently,
Eq.~(\ref{LKmodel}) reads $\dot{\mathbf{x}}=\mathbf{F}(\mathbf{x})$
and the linearization accordingly $\dot{\boldsymbol{\delta}
  \mathbf{x}}=\mathrm{D}F(\mathbf{x})\,\boldsymbol{\delta}
\mathbf{x}$. In this notation we introduce time-delayed feedback on a
single laser by
\begin{equation}
  \dot{\mathbf{x}}=\mathbf{F}(\mathbf{x})+\sigma\,\mathbf{H}(\mathbf{x}_{\tau}),
  \label{fullsystem}
\end{equation}
where
$\mathbf{H}(\mathbf{x})=\exp(-\mathrm{i}\,\omega_0\,\tau)(x_1,x_2,0)^\top$
describes linear coupling by reinjection of laser light into the
cavity. The carrier density is not directly affected by the
coupling. The coupling rate $\sigma$ denotes the feedback strength and
depends on the reflectivity of the distant mirror as well as absorbers
in the light path. The exponential term is a remainder of the
transformation from optical phase to relative phase. The linearization
of the laser equation with delayed feedback reads
\begin{equation}
  \dot{\boldsymbol{\delta} \mathbf{x}}=\mathrm{D}F(\mathbf{x})\,
  \boldsymbol{\delta} \mathbf{x}+\sigma\,\mathrm{D}H(\mathbf{x_{\tau}})\,\boldsymbol{\delta} \mathbf{x_{\tau}}
  \label{linsystem}
\end{equation}
and can be used to calculate LEs for a given trajetory
$\mathbf{x}(t)$, which is obtained by integration of
Eq.~(\ref{fullsystem}). The maximal LE $\lambda_{\mathrm{m}}$ is
defined by
\begin{equation}
  \lambda_{\mathrm{m}}=\lim_{t\rightarrow\infty}\frac{1}{t}\ln\left\{\frac{\|\boldsymbol{\delta} \mathbf{x}(t)\|}{\|\boldsymbol{\delta} \mathbf{x}(t_0)\|}\right\}
  \label{lyapcalc}
\end{equation}
The norm $\|\mathbf{a}\|$ in our case is $\sqrt{a_1^2+a_2^2}$, meaning
that we use only the electric field components of the Lyapunov
trajectory $\boldsymbol{\delta} \mathbf{x}(t)$. The choice of the norm
does not influence the final result, but may have strong influence on
the required length of the trajectory to calculate an exponent with
sufficiently high accuracy.

Since the system under investigation is a delay system, we formally
obtain infinitely many LEs depending on the initial conditions of the
linear system. In practice, for random initial conditions after some
transient time, the system relaxes onto the most unstable mode
revealing the maximal exponent. In the following, we concentrate on
the description and prediction of the dependence of
$\lambda_{\mathrm{m}}$ on the feedback parameter $\sigma$.

\section{Strong and weak chaos}

\subsection{The sub-Lyapunov exponent and its experimental
  measurement}
\label{subsec:subexp}

Conditional LEs play an important role for synchronization in networks
of coupled dynamical systems. More general, when a large system can be
divided into two subsystems forming a drive-response scheme (driver:
$\mathcal{D}$, response: $\mathcal{R}$), it makes sense to define a
conditional LE $\lambda_{\mathcal{R}}$ of the driven system
$\mathcal{R}$ in order to classify the dynamics or to predict
synchronization properties \cite{Pecora1990}. The conditional exponent
is often referred to as `sub-Lyapunov exponent' to stress the fact,
that it characterizes a subsystem. If the sub-exponent is negative,
$\mathcal{R}$ is in a state of generalized synchronization with
$\mathcal{D}$. Otherwise $\mathcal{R}$ is autonomously chaotic,
meaning that its state is not determined by the state of
$\mathcal{D}$.

The sign of a sub-LE can be measured experimentally. To this end a
copy $\mathcal{R}'$ of $\mathcal{R}$ has to be created. $\mathcal{R}'$
is then coupled to $\mathcal{D}$ in the same way that $\mathcal{R}$
is, meaning both units receive the same driving signal. If
$\lambda_{\mathcal{R}}<0$, then $\mathcal{R}'$ and $\mathcal{R}$
synchronize completely, otherwise not. This procedure is known as the
Abarbanel test \cite{Abarbanel1996} for generalized synchronization
between $\mathcal{D}$ and $\mathcal{R}$.

A similar situation is present in case of a system with time-delayed
feedback. Although quite against intuition, one can imagine
partitioning of the delay system into non-delayed dynamical unit
$\mathcal{A}\mathrel{\widehat{=}}\mathcal{R}$ and transmission line
$\mathcal{A_{\tau}}\mathrel{\widehat{=}}\mathcal{D}$. The sub-Lyapunov
exponent of the unit arises only from the instantaneous part of the
equations of motion, and has therefore been referred to as
`instantaneous Lyapunov exponent' \cite{Heiligenthal2011}. However, in
this paper we keep the notion `sub-Lyapunov exponent' and denote it as
$\lambda_0$. Formally, from the equations of motion
\begin{equation}
  \dot{\mathbf{x}}=\mathbf{F}(\mathbf{x})+\sigma\,\mathbf{H}(\mathbf{x_{\tau}})
  \label{delaysystem}
\end{equation}
we obtain $\lambda_0$ by integrating the linearization
\begin{equation}
  \dot{\boldsymbol{\delta} \mathbf{x}}_0=\mathrm{D}F(\mathbf{x})\,\boldsymbol{\delta} \mathbf{x_0}
  \label{linlambda0}
\end{equation}
\begin{equation}
  \lambda_0=\lim_{t\rightarrow\infty}\frac{1}{t}\ln\left\{\frac{\|\boldsymbol{\delta} \mathbf{x_0}(t)\|}{\|\boldsymbol{\delta} \mathbf{x_0}(t_0)\|}\right\}.
  \label{lambda0def}
\end{equation}
Alternatively, we can define $\lambda_0$ via the evolution operator of
Eq.~\ref{linlambda0}
\begin{equation}
  \boldsymbol{\delta} \mathbf{x_0}(t)=U(t,t_0)\,\boldsymbol{\delta} \mathbf{x_0}(t_0),
\end{equation}
with $U(t_0,t_0)=\mathbb{1}$. Then
\begin{equation}
  \lambda_0=\max_i\lim_{t\rightarrow\infty}\frac{1}{2\,t}\ln\!\left(\eig_i\{U^\top(t,t_0)\,U(t,t_0)\}\right).
\end{equation}
It is important to mention, that although the delay term is skipped in
the defining equation for $\lambda_0$, the delay parameters $\sigma$
and $\tau$ enter indirectly via the trajectory $\mathbf{x}(t)$, which
is inserted in the linearization Eq.~(\ref{linlambda0}). As we show
later, this dependence is non-negligible and it is possible to change
the sign of $\lambda_0$ only by variation of a delay parameter.

The consequence of the sign of $\lambda_0$ is documented in recent
work \cite{Heiligenthal2011}. If $\lambda_0>0$, we call the resulting
chaotic dynamics `strong chaos', else if $\lambda_0<0$ but the delay
system is still chaotic, we define `weak chaos'. The sign of
$\lambda_0$ has two major implications. Firstly, for a single delay
system it determines the scaling of the maximal LE with delay
time. Secondly, for a network of delay-coupled units it determines the
possibility to synchronize the units at large delay times. If the
units are strongly chaotic, the delayed coupling cannot compensate for
the exponential divergence of two nearby trajectories of any two
systems in the network, and therefore synchronization is not
possible. On the other hand, for weak chaos, synchronization is in
principle possible and depends on the network topology.

\begin{figure}
  \includegraphics[height=0.4\columnwidth]{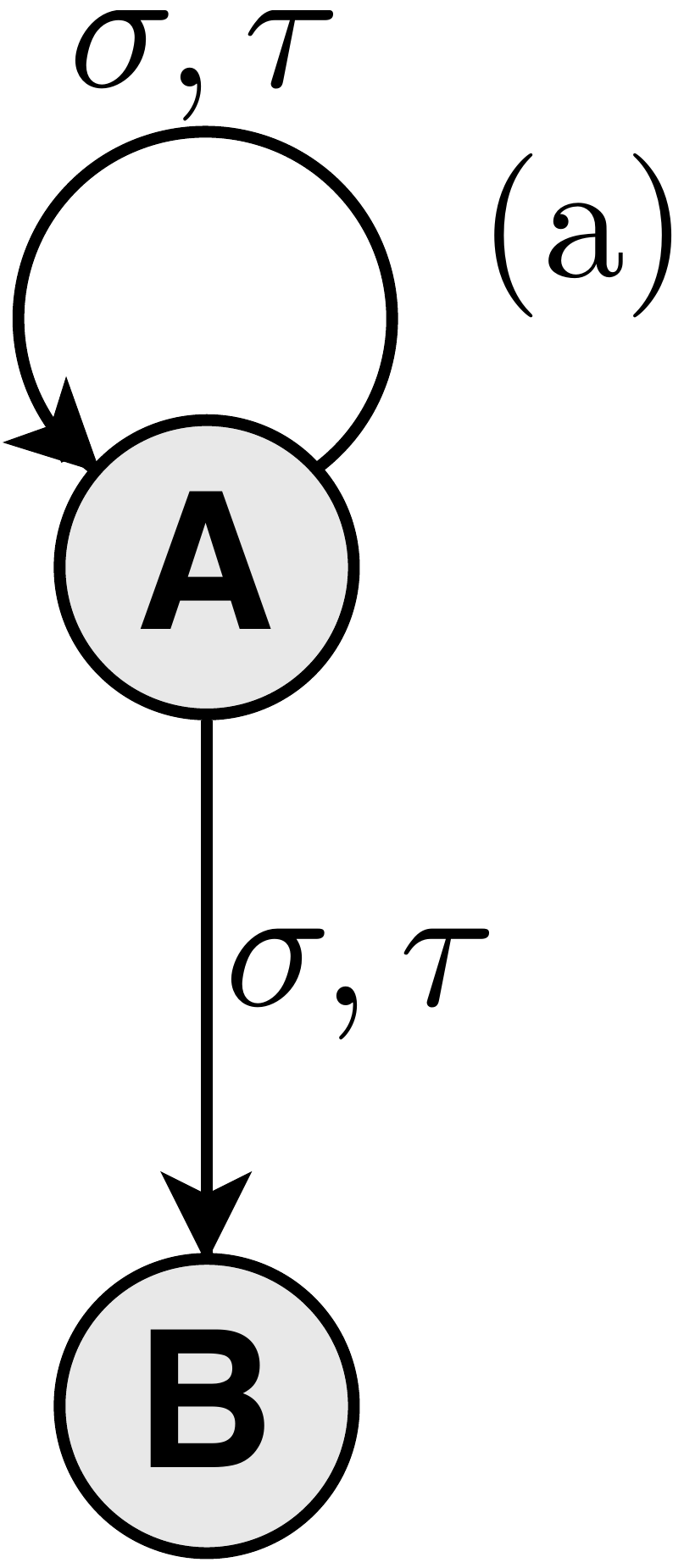}
  \hspace*{0.5cm} \includegraphics[height=0.4\columnwidth]{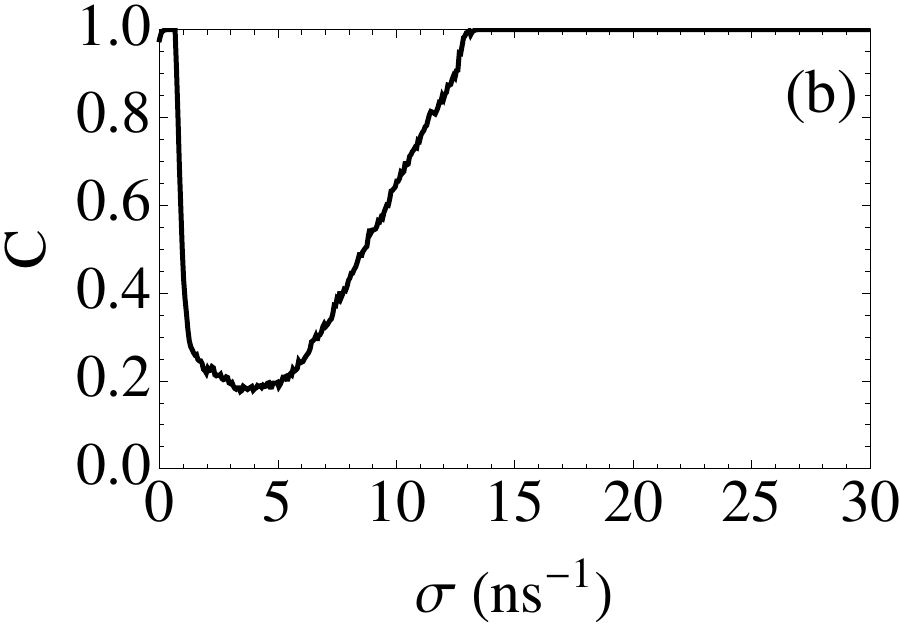}
  \caption{(a)~Configuration to measure the difference between strong
    and weak chaos. (b)~Simulated isochronal cross-correlation $C$
    between two lasers coupled according to the setup presented in (a)
    for a delay time $\tau = 10 \, \mathrm{ns}$ and a pump current $p
    = 1.02$ vs coupling strength $\sigma$.}
  \label{F_III_A_1}
\end{figure}

In analogy to the general Abarbanel test, we can measure the sign of
the sub-LE $\lambda_0$ using an auxiliary system approach as shown in
Fig.~\ref{F_III_A_1}(a). The system
$\mathcal{B}\mathrel{\widehat{=}}\mathcal{R}'$ is a copy of the
undelayed system $\mathcal{A}\mathrel{\widehat{=}}\mathcal{R}$. In the
presented coupling topology both units are receiving the same signal
$\mathcal{A_{\tau}}\mathrel{\widehat{=}}\mathcal{D}$. If we analyze
the equations of motion for this setup in the vicinity of the complete
synchronization manifold (SM) of $\mathcal{A}$ and $\mathcal{B}$, we
obtain in linear approximation the same system, by which we also have
defined $\lambda_0$. This means, that in the test setup both units
synchronize if and only if they are in weak chaos. Synchronization is
unstable if and only if they are in strong chaos. The result of a
numerical simulation of this setup for two coupled semiconductor
lasers is shown in Fig.~\ref{F_III_A_1}(b) showing the
cross-correlation between $\mathcal{A}$ and $\mathcal{B}$ in
dependence on the coupling strength $\sigma$. A cross-correlation
$C=1$ corresponds to complete synchronization and implies the presence
of weak chaos. The drop-down of the correlation indicates transitions
from weak to strong and back to weak chaos by increasing $\sigma$. We
discuss this behavior in detail in Sec.~\ref{sec:sfblaser}.

\subsection{Scaling of the maximal Lyapunov exponent with the delay
  time}

As we show in the following, the sub-LE $\lambda_0$, as introduced by
Eqs.~\eqref{delaysystem}-\eqref{lambda0def}, determines the scaling
behavior of the maximal LE $\lambda_{\mathrm{m}}$, as introduced by
Eqs.~\eqref{fullsystem}-\eqref{lyapcalc}, with the delay time.  Note
that there is always a spectrum of LEs, in this case a countable
infinite set. $\lambda_{\mathrm{m}}$ is the maximal exponent of this
spectrum. Further we assume the existence of a single attractor and
the ergodicity of the trajectory $\mathbf{x}(t)$ so that we can also
skip the dependence on the initial conditions.

We now distinguish between two cases: a) Strong chaos,
$\lambda_0>0$. b) Weak chaos, $\lambda_0<0$. In order to derive a
scaling relation $\lambda_{\mathrm{m}}(\tau)$, we have to assume, that
$\lambda_0(\tau)=\const$. In a strict sense this is never true, but
for sufficiently large $\tau$ one always observes a saturation effect
and a decrease of remaining fluctuations in $\lambda_0(\tau)$, so that
the assumption of a constant value is valid.

\paragraph{Strong chaos, $\lambda_0>0$.}

We start from the linearization Eq.~(\ref{linsystem}). If
$\lambda_0>0$, for large $\tau$ the instantaneous term becomes
dominant and the delay term becomes negligible. This can be seen from
the coordinate transformation
\begin{equation}
  \boldsymbol{\delta} \mathbf{x}(t)=\e^{\lambda_0\,t}\,\boldsymbol{\delta} \mathbf{z}(t),
\end{equation}
which results in
\begin{equation}
  \dot{\boldsymbol{\delta} \mathbf{z}}=[\mathrm{D}F(\mathbf{x})-\lambda_0\cdot\mathbb{1}]\,
  \boldsymbol{\delta} \mathbf{z}+\sigma\e^{-\lambda_0\,\tau}\mathrm{D}H(\mathbf{x_{\tau}})\,\boldsymbol{\delta} \mathbf{z_{\tau}}.
\end{equation}
For large delay times the delay term in this equation becomes
exponentially small and can be neglected. The resulting Lyapunov
exponent of the transformed system can therefore be estimated as
$\lambda_{\mathbf{z}}\approx 0$. This implies that in the original
coordinates $\lambda_{\mathrm{m}}\approx\lambda_0$, meaning that the
Lyapunov becomes independent of $\tau$ for large delays.

\paragraph{Weak chaos, $\lambda_0<0$.}

In this case we can estimate the scaling of the LE by considering a
stroboscopic sequence of the evolution of our linear system,
$\boldsymbol{\delta} \mathbf{x_n}(\theta)=\boldsymbol{\delta}
\mathbf{x}(\theta+n\,\tau)$ with $n\in\mathbb{Z}$ and
$\theta\in]-\tau,0]$. One can introduce a growth factor (Lyapunov
multiplier) by $\|\boldsymbol{\delta}
\mathbf{x_{n+1}}\|=\mu_n\,\|\boldsymbol{\delta} \mathbf{x_n}\|$, so
that the LE becomes
\begin{equation}
  \lambda_{\mathrm{m}}=\lim_{l\rightarrow\infty}\frac{1}{l\,\tau}\sum_{n=1}^{l}\ln \mu_n=\frac{1}{\tau}\ln C,
\end{equation}
where $C$ is the geometric mean of all multipliers. In the following
we show, that for sufficiently large delay times the multipliers
$\mu_n$ do not depend on $\tau$, hence the LE is of order $\tau^{-1}$,
meaning that in weak chaos a perturbation of the chaotic system grows
on the timescale of the delay time. To this end, we introduce the
variation-of-constants formula, which provides an integral version of
our initial delay differential equation~(\ref{linsystem}), and
evaluate it at $\theta=0$
\begin{equation}
  \begin{split}
    \boldsymbol{\delta} \mathbf{x_{n+1}}(0)=&\; U_n(0,-\tau)\,\boldsymbol{\delta} \mathbf{x_n}(0)\\
    &+\sigma\!\int\limits_{-\tau}^0
    \mathrm{d}t\,U_n(0,t)\,\mathrm{D}H[\mathbf{x_n}(t)]\,\boldsymbol{\delta}
    \mathbf{x_n}(t).
  \end{split}
  \label{vocformula}
\end{equation}
It contains the evolution operator $U_n(t_2,t_1)$ of the auxiliary
system Eq.~(\ref{linlambda0}) on the $n$-th $\tau$-interval. This
operator is exponential in $t_2-t_1$, i.\,e. with respect to a
suitable matrix norm it holds that for $t_2>t_1$
\begin{equation}
  \|U(t_2,t_1)\|<U_0\,\exp[\alpha\,(t_2-t_1)]
  \label{uest}
\end{equation}
with $\lambda_0<\alpha<0$. The bound provided by $U_0$ and $\alpha$
should cover potential bursts typical for the linearization of a
chaotic flow. Because of this exponential bound, there is some time
$\tau_0\propto\alpha^{-1}$, such that the term with $U(0,-\tau)$ can
be neglected in Eq.~(\ref{vocformula}), if
$\tau>\tau_0$. Additionally, the integral has only significant
contributions in a small range close to the end of the integration
interval, namely for $t\in[-\tau_0,0]$. This means that a further
increase of $\tau$ beyond $\tau_0$ does not affect the
integral. Therefore the multiplier introduced above can be estimated
by
\begin{equation}
  \begin{split}
    \mu_n=&\;\frac{\sigma}{\|\boldsymbol{\delta}
      \mathbf{x_n}(0)\|}\left\|\int\limits_{-\tau_0}^0
      \mathrm{d}t\,U_n(0,t)\,
      \mathrm{D}H[\mathbf{x_n}(t)]\,\boldsymbol{\delta} \mathbf{x_n}(t)\right\|\\
    &+\mathcal{O}\left(\e^{-\alpha\,\tau}\right).
  \end{split}
\end{equation}
In leading order, this expression does not depend on the delay
time. It depends on $\lambda_0$, $\sigma$ and a set $\mathbf{q}$ of
other (yet unknown) statistical properties of the chaotic
trajectory. Hence we can write
\begin{equation}
  \lambda_{\mathrm{m}}=\frac{1}{\tau}\ln C(\lambda_0,\sigma,\mathbf{q}).
\end{equation}
For comparison: Given a Floquet problem, in which the driving
trajectory is $\tau$-periodic ($\mathbf{x}(t)=\mathbf{x}(t+\tau)$) and
the coupling is linear and diagonal ($\mathrm{D}H[\mathbf{x}(t)]
\equiv \mathbb{1}$), we obtain $C=-\sigma/\lambda_0$.

\section{Single laser with time-delayed self-feedback}
\label{sec:sfblaser}

\subsection{Transitions between strong and weak chaos}

We now discuss the most simple system which can show strong and weak
chaos: a single laser with time-delayed self-feedback. The laser
dynamics is modeled by the LK equations. The parameter values can be
found in the appendix. If not mentioned differently, all diagrams are
made for a delay time of $\tau = 10 \, \mathrm{ns}$ and a pump current
of $p = 1.02$.

\begin{figure}
  \includegraphics[height=0.35\columnwidth]{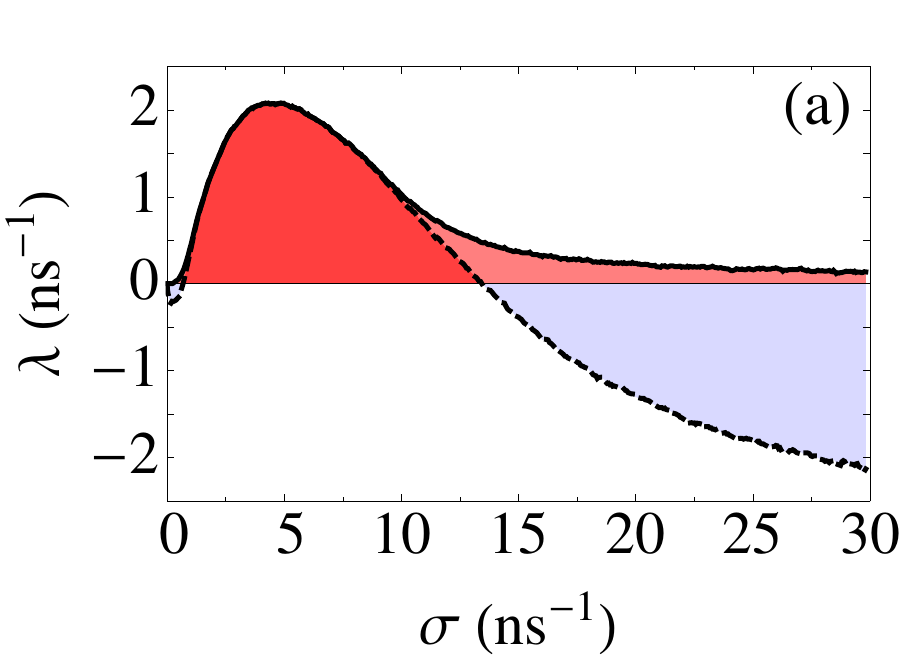}
  \hspace*{\fill} \includegraphics[height=0.35\columnwidth]{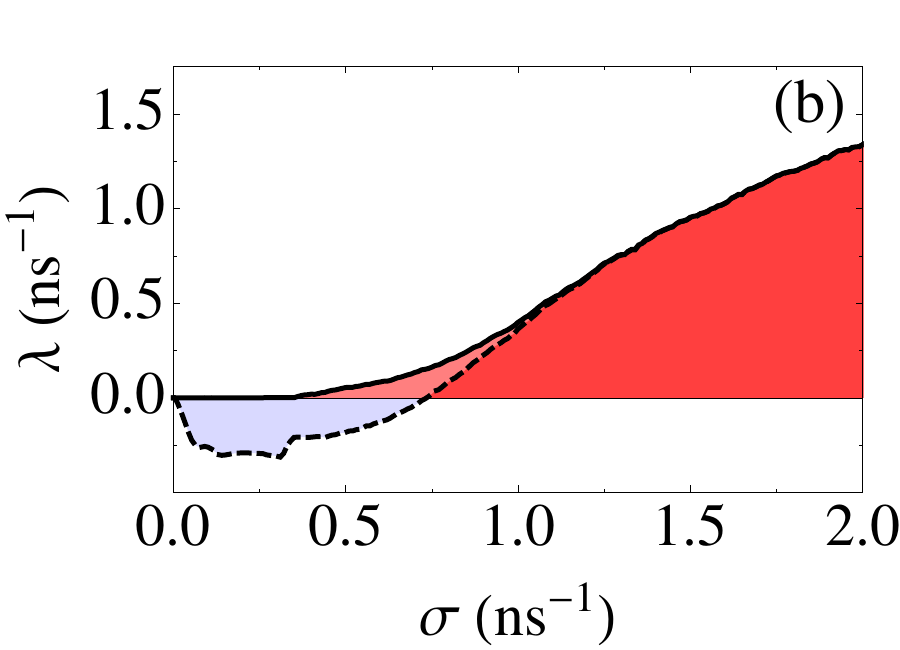}
  \caption{(Color online) (a)~Maximal Lyapunov exponent (LE)
    $\lambda_{\mathrm{m}}$ (solid line) and sub-LE $\lambda_{0}$
    (dashed line) of a single laser with self-feedback for a delay
    time $\tau=10\,\mathrm{ns}$ vs coupling strength $\sigma$.
    (b)~Enlarged view of left diagram for small coupling strengths
    $\sigma$.}
  \label{F_IV_A_1}
\end{figure}

Fig.~\ref{F_IV_A_1}(a) shows the maximal LE $\lambda_{\mathrm{m}}$
(solid line) and sub-LE $\lambda_{0}$ (dashed line) of a single laser
with self-feedback in dependence on the coupling strength
$\sigma$. Although the coupling strength does not enter directly into
the conditional equation of the sub-LE, as it would be the case
e.\,g. for a linear damping term with $\sigma$, the sign of the sub-LE
changes from positive to negative for large coupling strengths only
through the different dynamics $\mathbf{x}(t)$ which is inserted into
Eq.~\eqref{linlambda0}. Thus, there is a transition from strong to
weak chaos with growing $\sigma$. In the region of small coupling
strenghts, Fig.~\ref{F_IV_A_1}(b), however, there is an additional
region of weak chaos where the laser is already chaotic,
i.\,e. $\lambda_{\mathrm{m}} > 0$, but the sub-LE is still
negative. Hence, there is an additional transition from weak to strong
chaos upon increasing $\sigma$ for small coupling strengths.

Overall, we observe a transition from periodic behavior (Goldstone
mode with $\lambda_{\mathrm{m}}=0$) to weak chaos and from there
transitions to strong chaos and back to weak chaos upon monotonically
increasing the coupling strength.

\begin{figure}
  \includegraphics[height=0.31\columnwidth]{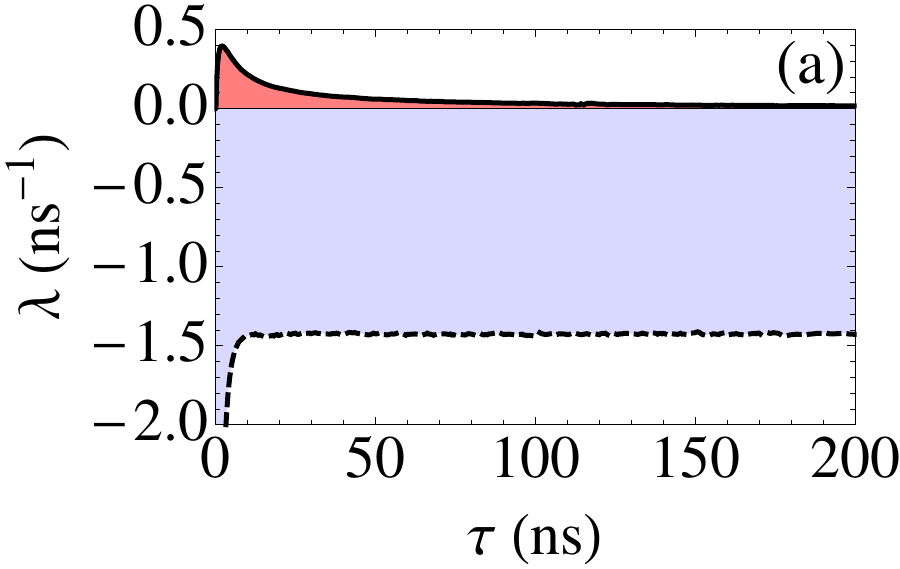}
  \hspace*{\fill}
  \includegraphics[height=0.30\columnwidth]{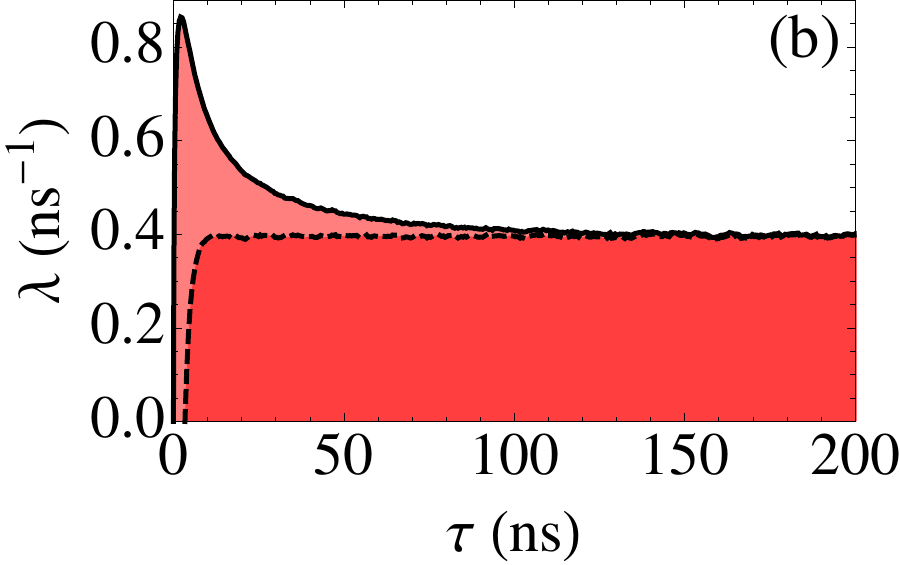}
  \newline \hspace*{3mm}
  \includegraphics[height=0.33\columnwidth]{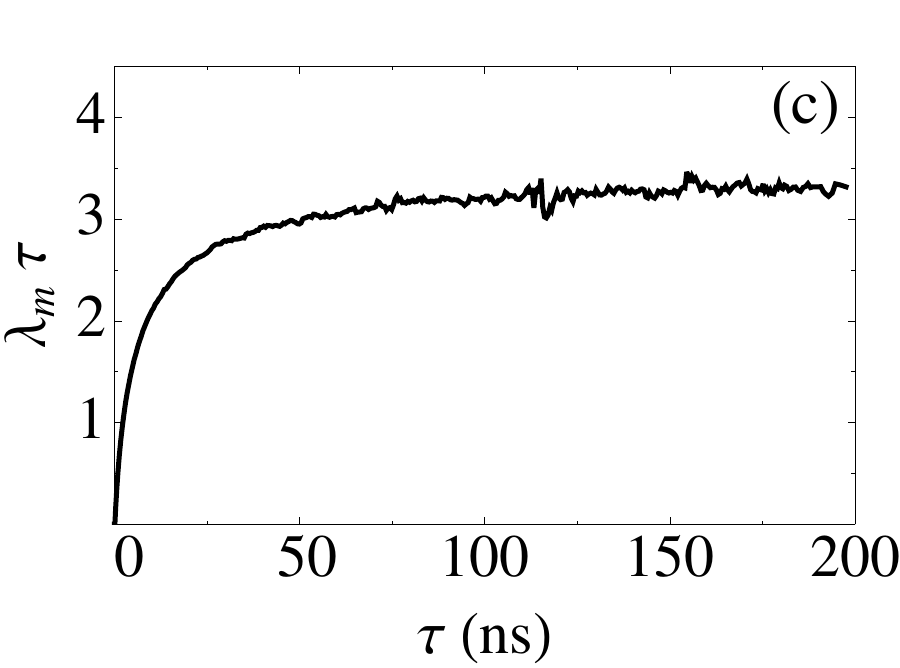}
  \hspace*{\fill}
  \includegraphics[height=0.30\columnwidth]{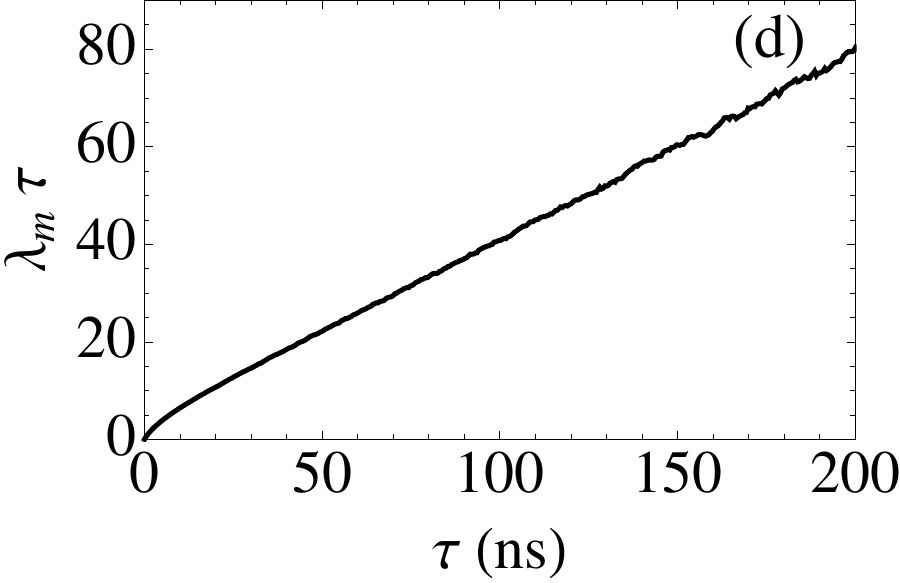}
  \newline \hspace*{\fill}
  \includegraphics[height=0.34\columnwidth]{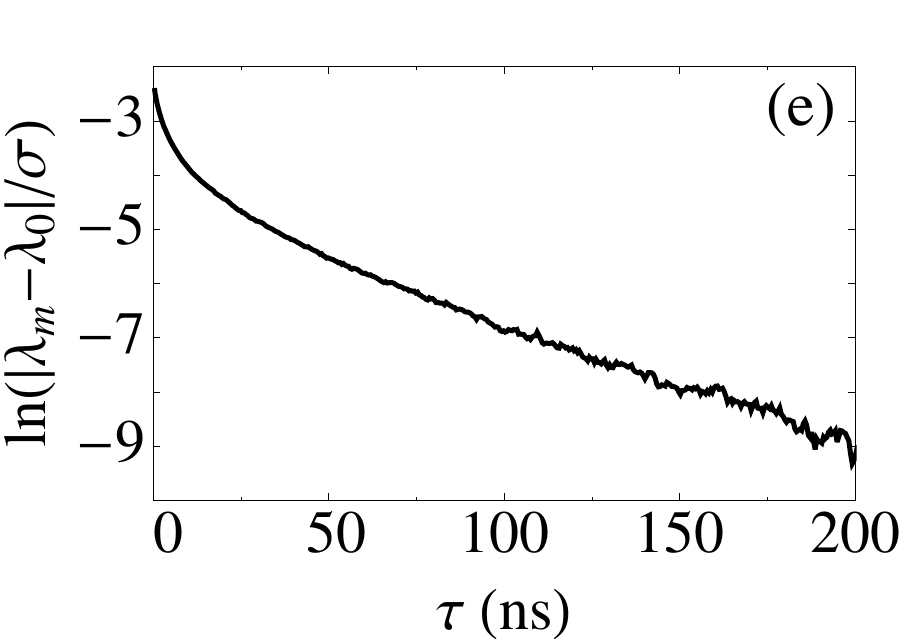}
  \caption{(Color online) Maximal LE $\lambda_{\mathrm{m}}$ (solid
    line) and sub-LE $\lambda_{0}$ (dashed line) of a single laser
    with self-feedback for (a) weak chaos ($\sigma = 21 \,
    \mathrm{ns}^{-1}$) and (b) strong chaos ($\sigma = 12 \,
    \mathrm{ns}^{-1}$) vs delay time $\tau$. Product
    $\lambda_{\mathrm{m}} \, \tau$ of a single laser with
    self-feedback for (c) weak chaos ($\sigma = 21 \,
    \mathrm{ns}^{-1}$) and (d) strong chaos ($\sigma = 12 \,
    \mathrm{ns}^{-1}$) vs delay time $\tau$. (e)
    $\ln(|\lambda_{\mathrm{m}}-\lambda_{0}|/\sigma)$ of a single laser
    with self-feedback vs delay time $\tau$ for strong chaos ($\sigma
    = 12 \, \mathrm{ns}^{-1}$).}
  \label{F_IV_A_2}
\end{figure}

Fig.~\ref{F_IV_A_2} shows the different behaviors of the maximal LE
and the sub-LE for the regimes of weak, Fig.~\ref{F_IV_A_2}(a), and
strong chaos, Fig.~\ref{F_IV_A_2}(b), in dependence on the delay time
$\tau$. As soon as the delay time is large compared with the internal
time scales of the laser, the sub-LE $\lambda_0$ remains constant with
increasing $\tau$ for both weak and strong chaos. For weak chaos,
$\lambda_{\mathrm{m}}$ decreases like 1/$\tau$ for growing $\tau$. For
strong chaos, $\lambda_{\mathrm{m}}$ converges exponentially to the
positive sub-LE $\lambda_0$.

The product $\lambda_{\mathrm{m}} \, \tau$ is the relevant
dimensionless quantity describing chaoticity in systems with
delay. For weak chaos, this product saturates at a constant value
[Fig.~\ref{F_IV_A_2}(c)] which depends on the coupling strength. This
dependence is caused by different levels of fluctuations for different
$\sigma$. Also the delay time $\tau$ at which the product
$\lambda_{\mathrm{m}} \, \tau$ has reached the saturation value up to
a fixed distance $\delta$ depends on $\sigma$: The closer one is at
the critical coupling strength $\sigma_{\mathrm{crit}}$ where
$\lambda_0 = 0$, the larger the delay time has to be,
i.\,e. saturation happens later for larger $\tau$.

For strong chaos, on the contrary, $\lambda_{\mathrm{m}} \, \tau$
grows linearly with increasing $\tau$ [Fig.~\ref{F_IV_A_2}(d)] since
$\lambda_{\mathrm{m}}$ gets constant as shown in
Fig.~\ref{F_IV_A_2}(b). Fig.~\ref{F_IV_A_2}(e) confirms that the
convergence $\lambda_{\mathrm{m}} \to \lambda_0 > 0$ happens
exponentially with $\tau$. The characteristic exponent of this
convergence, however, is much larger (i.\,e. the convergence is
slower) than one would expect from analytic calculations for a simple
model with constant coefficients. This effect is caused by the
fluctuations of $\mathrm{D}F[\mathbf{x}(t)]$ which act like
multiplicative noise in Eq.~\eqref{linsystem} and \eqref{linlambda0}
\cite{Juengling2012}.

\begin{figure}
  \includegraphics[height=0.30\columnwidth]{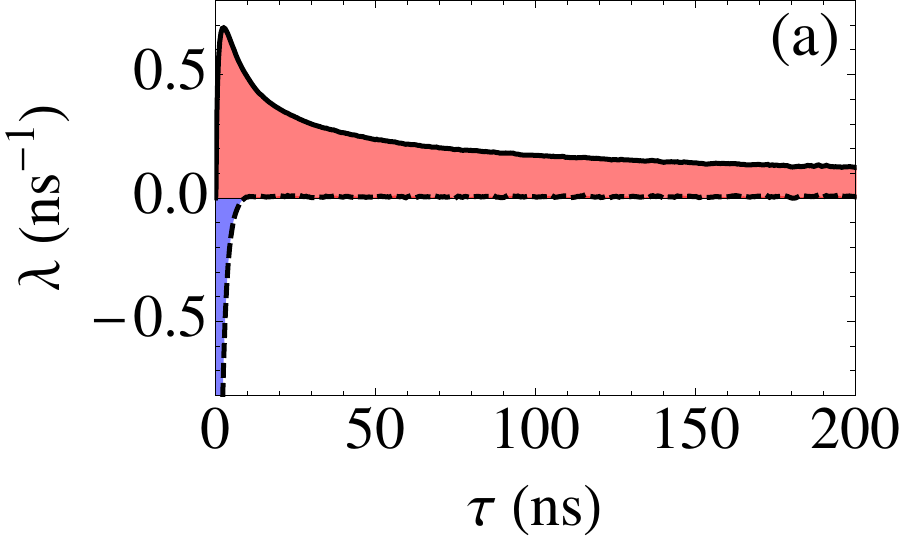}
  \hspace*{\fill}
  \includegraphics[height=0.32\columnwidth]{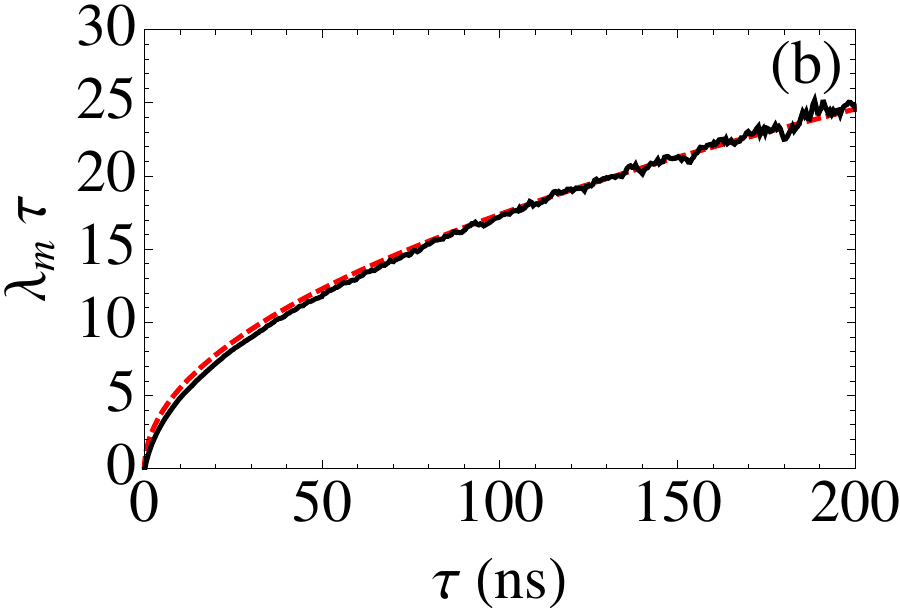}
  \caption{(Color online) (a) Maximal LE $\lambda_{\mathrm{m}}$ (solid
    line) and sub-LE $\lambda_{0}$ (dashed line) and (b) product
    $\lambda_{\mathrm{m}} \, \tau$ (black solid line) of a single
    laser with self-feedback in comparison with the least-squares fit
    of $\sqrt{\tau}$ (red dashed line) for critical coupling strength
    where transition between strong and weak chaos occurs
    ($\sigma_{\mathrm{crit}} = 13.4 \, \mathrm{ns}^{-1}$) vs delay
    time $\tau$.}
  \label{F_IV_A_3}
\end{figure}

Fig.~\ref{F_IV_A_3} shows the special limit case between strong and
weak chaos when the sub-LE $\lambda_0 = 0$ for suffiently large delay
times $\tau$. In Fig.~\ref{F_IV_A_3}(a) one can see that the maximal
LE still decays with increasing $\tau$. However, it does so very
slowly. Fig.~\ref{F_IV_A_3}(b) shows the consequence for the product
$\lambda_{\mathrm{m}} \, \tau$ at the critical point: Neither does it
grow linearly with $\tau$ like for strong chaos, nor does it saturate
at a constant value for finite delay times like for weak chaos. The
point where the product $\lambda_{\mathrm{m}} \, \tau$ saturates is
only reached for infinitely large delay times $\tau \to
\infty$. $\lambda_{\mathrm{m}} \, \tau$ rather grows like
$\sqrt{\tau}$ as is shown by the least-squares fit with $\sqrt{\tau}$
(red dashed line) in Fig.~\ref{F_IV_A_3}(b). It turns out that this
behavior can be explained as well with the effect of the
multiplicative fluctuations of $\mathrm{D}F[\mathbf{x}(t)]$. These
fluctuations can be shown to appear in simple systems in which they
are modeled as multiplicative white noise \cite{Juengling2012}.

\begin{figure}
  \includegraphics[width=0.55\columnwidth]{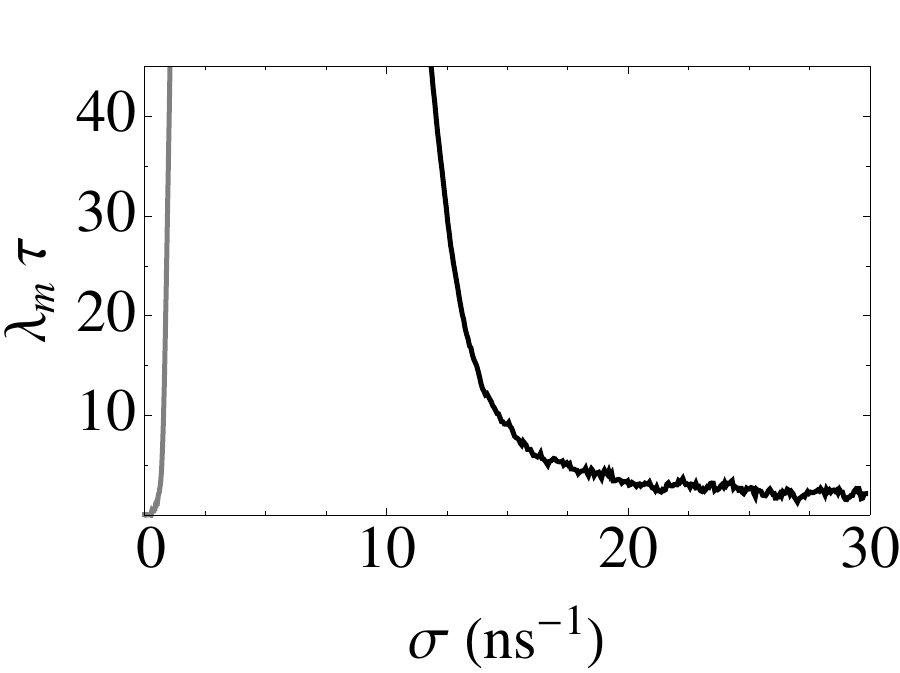}
  \caption{Product $\lambda_{\mathrm{m}} \, \tau$ of a single laser
    with self-feedback vs coupling strength $\sigma$ for a delay time
    $\tau=100\,\mathrm{ns}$.}
  \label{F_IV_A_4}
\end{figure}

Fig.~\ref{F_IV_A_4}(a) shows that for weak chaos and a very large
delay time $\tau \to \infty$, the product $\lambda_{\mathrm{m}} \,
\tau$ undergoes a phase transition and diverges in the proximity of
the critical coupling strengths $\sigma_{\mathrm{crit},1}$ (gray line)
and $\sigma_{\mathrm{crit},2}$ (black line).

We were able to find the occurence of strong and weak chaos and the
transitions between them by changing the coupling strength $\sigma$
also for the R\"ossler and Lorenz dynamics. The Stuart-Landau,
FitzHugh-Nagumo and continuous Ikeda dynamics only show weak chaos.

\subsection{Scaling of the sub-Lyapunov exponent with the laser pump
  current}

\begin{figure}
  \hspace*{1cm}
  \includegraphics[width=0.55\columnwidth]{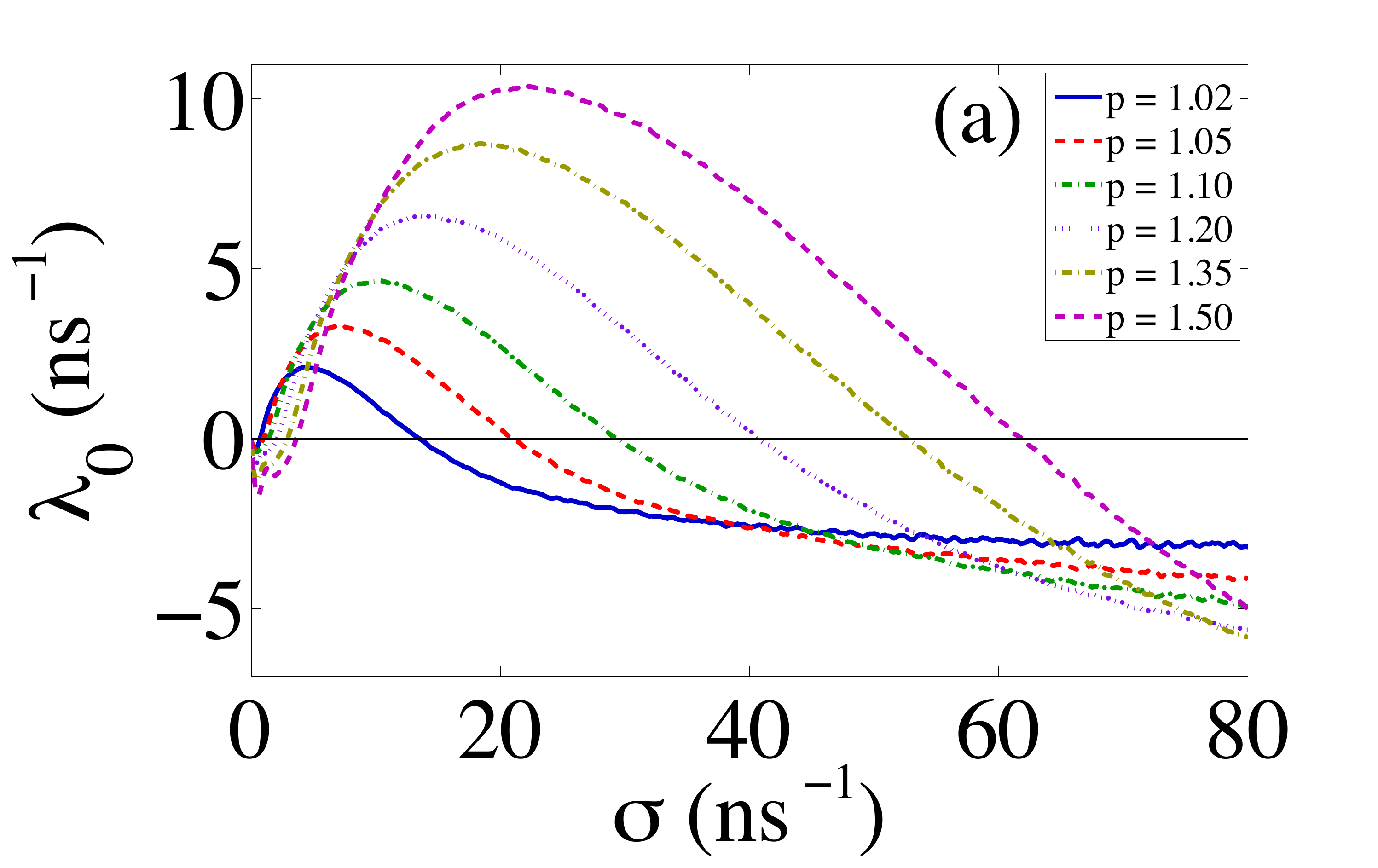}
  \newline \hspace*{1cm}
  \includegraphics[width=0.55\columnwidth]{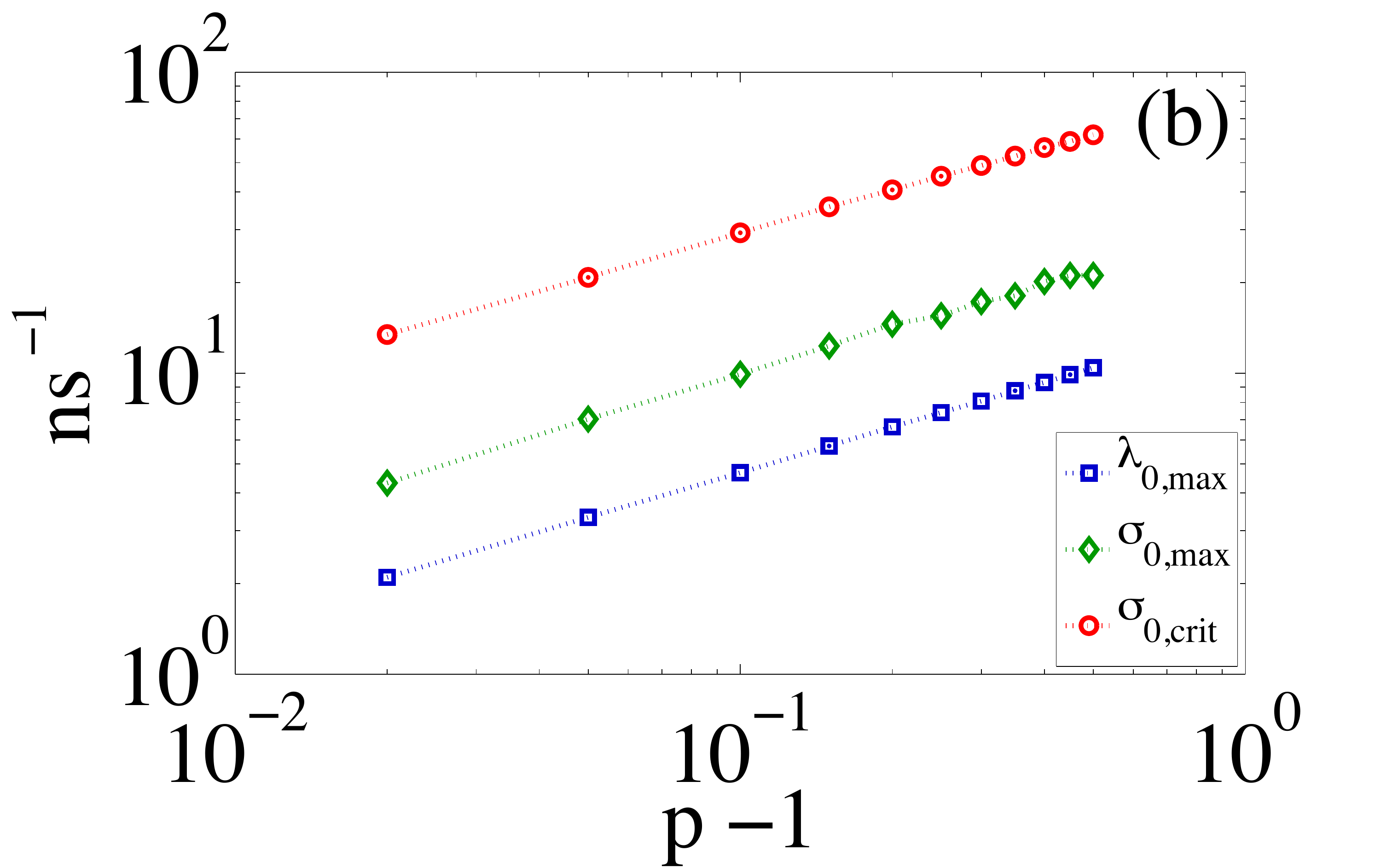}
  \newline
  \includegraphics[width=0.55\columnwidth]{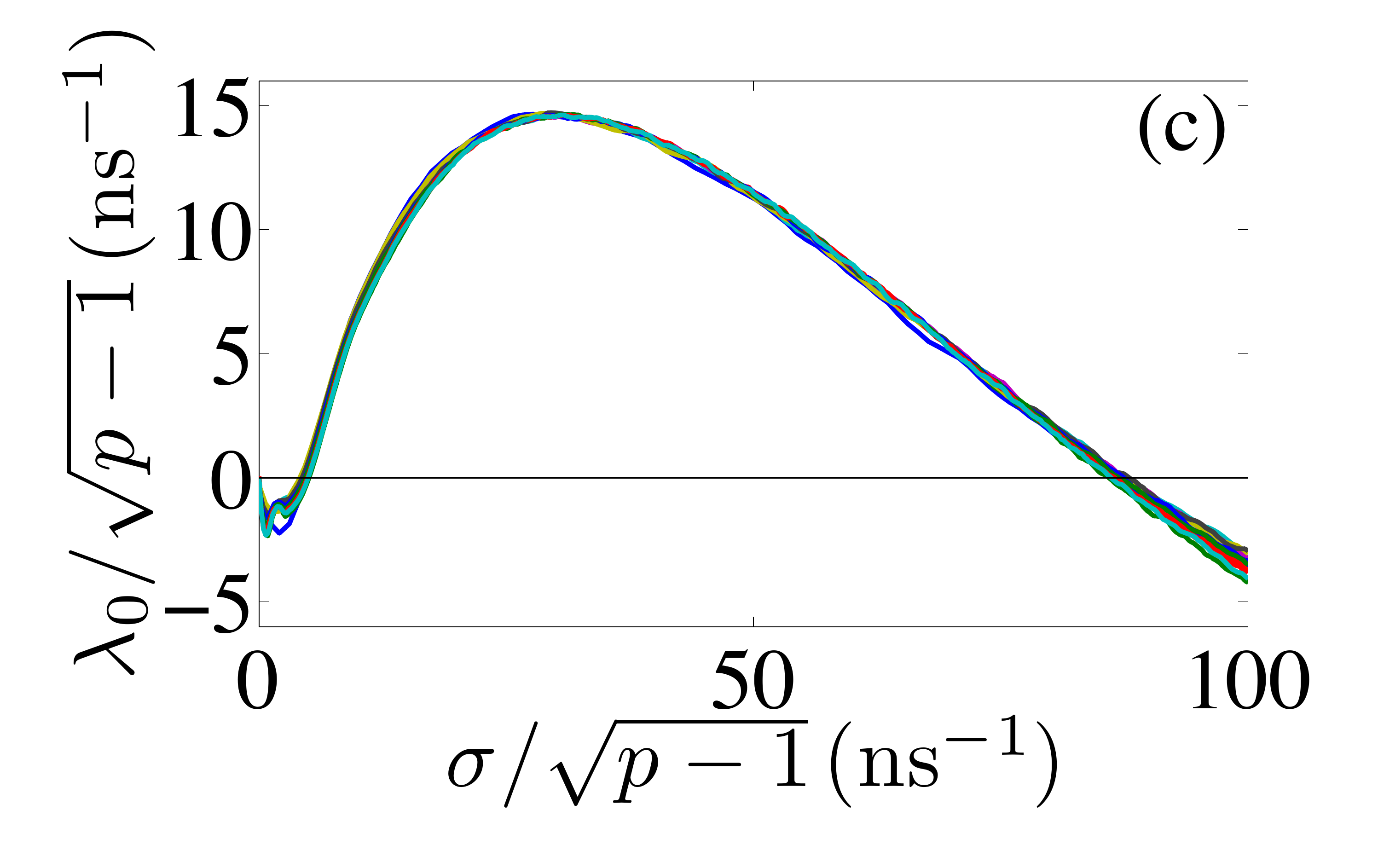}
  \caption{(Color online) (a)~Sub-LE $\lambda_{0}$ of a single laser
    with self-feedback for a delay time $\tau=10\,\mathrm{ns}$ and
    several different pump currents $p$ vs coupling strength
    $\sigma$. (b)~Maxima $\lambda_{0,\mathrm{max}}$ of the sub-LEs
    $\lambda_{0}$, coupling strengths $\sigma_{0,\mathrm{max}}$ of
    these maxima, and critical coupling strengths
    $\sigma_{\mathrm{crit}}$ where $\lambda_0 = 0$ for several
    different pump currents $p$ vs effective pump current $p-1$ in
    double-logarithmic scale. (c)~Data collapse of rescaled sub-LEs
    $\lambda_0/\sqrt{p-1}$ vs rescaled coupling strengths
    $\sigma/\sqrt{p-1}$ for eleven different pump currents $p$ ranging
    from $p = 1.02$ to $p = 1.50$.}
  \label{F_IV_B_1}
\end{figure}

We study the dependence of the sub-LE on the laser pump current.
Fig.~\ref{F_IV_B_1}(a) shows the sub-LE as a function of the feedback
strength $\sigma$ for different values of the pump current.  We find
that the curves all follow the same pattern described in the previous
section. The sub-LE is negative for small and for large values of the
feedback strength.  The agreement is even quantitative. We find that
the sub-LE and coupling strength scale with the square root of the
effective pump current $\sqrt{p-1}$ above the lasing
threshold. Fig.~\ref{F_IV_B_1}(b) shows the maximum of the sub-LE
$\lambda_{0,\mathrm{max}}$, the feedback strength
$\sigma_{0,\mathrm{max}}$ for which the sub-LE is maximal, and the
critical feedback strength $\sigma_{0,\mathrm{crit}}$ where $\lambda_0
= 0$ as a function of the effective pump current $p-1$.  The data is
shown in a double-logarithmic scale, indicating a slope of $1/2$. In
Fig.~\ref{F_IV_B_1}(c) the sub-LEs are rescaled with $\sqrt{p-1}$. We
find that this scaling relation holds well for small coupling
strengths. For larger coupling strengths the sub-LEs diverge for
different pump currents.

To explain this scaling behavior we introduce the following rescaling
of the laser parameters
\begin{eqnarray*}
  N & = & \sqrt{\frac{G_{\mathrm{N}}}{\Gamma \, N_{\mathrm{sol}}}}\,\frac{n}{\sqrt{p-1}}\\
  E & = & \sqrt{\frac{\Gamma}{\gamma \, N_{\mathrm{sol}}}}\,\frac{\mathcal{E}}{\sqrt{p-1}}\\
  K & = & \frac{1}{\sqrt{G_{\mathrm{N}} \, N_{\mathrm{sol}}}}\,\frac{\sigma}{\sqrt{p-1}}\\
  s & = & \sqrt{\Gamma \, G_{\mathrm{N}} \, N_{\mathrm{sol}}}\,\sqrt{p-1}\,t.
\end{eqnarray*}
\noindent Such a scaling reproduces the scaling behavior that we found
numerically: The coupling strength $K$ and the Lyapunov exponent
(which scales inversely with time) scale with $\sqrt{p-1}$.  The LK
equations, Eqs.~\eqref{LKmodel}, can then be rewritten as
\begin{eqnarray}
  \frac{dE}{ds} & = & \frac{1+\mathrm{i}\,\alpha}{2} \, N \,E+K \, \mathrm{e}^{\mathrm{i} \,\theta} \, E(s-\sqrt{p-1}\,\tau)\nonumber\\
  \frac{dN}{ds} & = & \frac{\gamma}{\Gamma}\left[1-\frac{N}{c \, \sqrt{p-1}}-(c \, \sqrt{p-1} \, N + 1)|E|^2\right],\label{eq:rescaled}
\end{eqnarray}
\noindent with $c=\sqrt{G_{\mathrm{N}} N_{\mathrm{sol}}/\Gamma}$.  We
assume small coupling strength $K\ll 1$, small carrier densities
$N=\mathcal{O}(K)$, and a reasonably high pump current $c \,
\sqrt{p-1}=\mathcal{O}(1)$. The ratio of photon and carrier life
times, $\gamma/\Gamma$, is a small parameter as well. In leading order
we obtain the following equations:
\begin{eqnarray}
  \frac{dE}{ds} & = & \frac{1+\mathrm{i}\,\alpha}{2} \,N \, E + K \mathrm{e}^{\mathrm{i} \, \theta} E(s - c \, \sqrt{p-1} \, \tau)\nonumber\\
  \frac{dN}{ds} & = & \frac{\gamma}{\Gamma}\left(1-|E|^2\right).\label{eq:rescaledb}
\end{eqnarray}
These equations only depend on the pump current in the value of the
time delay. Since the exact value of the time delay does not influence
the sub-LE, we recover the scaling behavior found numerically. For
increasing coupling strength $K$ the mapping becomes less exact as can
be seen in Fig.~\ref{F_IV_B_1}(c). This results from the fact that the
rescaled model Eq.~\eqref{eq:rescaledb} is a weak coupling
approximation.

\subsection{Auto-correlations and space-time patterns for strong and
  weak chaos}

\begin{figure}
  \includegraphics[height=0.31\columnwidth]{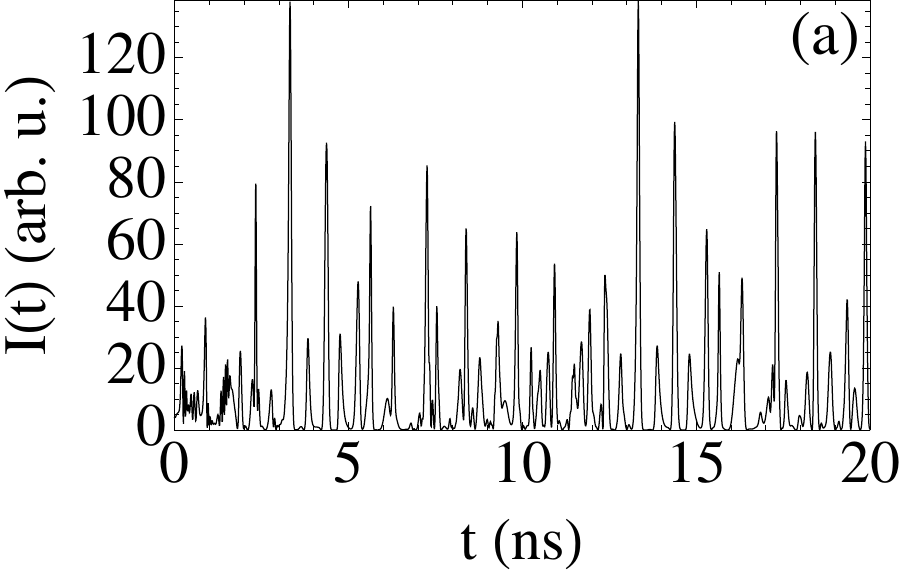}
  \hspace*{\fill} \includegraphics[height=0.31\columnwidth]{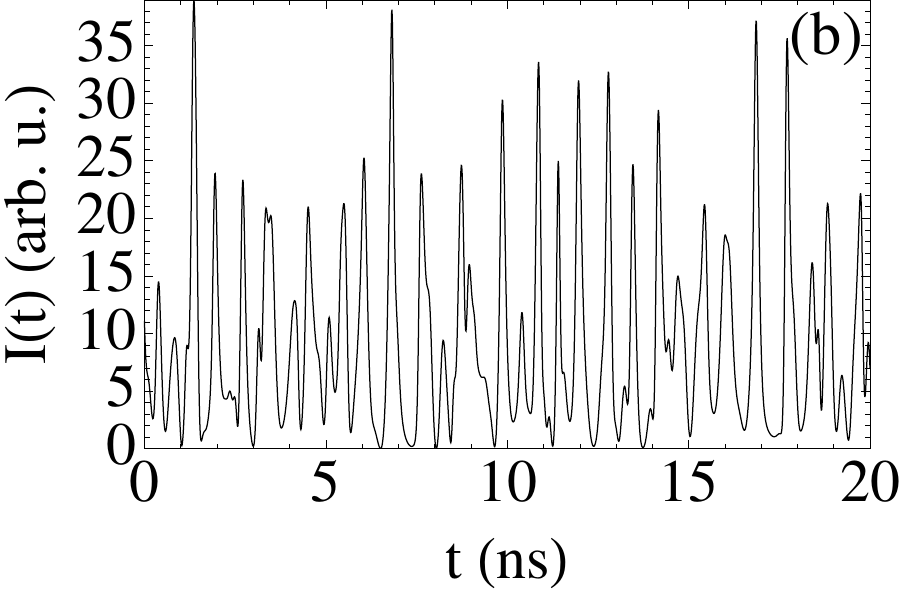}
  \caption{Example trajectories (intensity traces) of a single laser
    with a self-feedback of $\tau=10\,\mathrm{ns}$ for (a) weak chaos
    ($\sigma = 21 \, \mathrm{ns}^{-1}$) and (b) strong chaos ($\sigma
    = 6 \, \mathrm{ns}^{-1}$) vs time $t$.}
  \label{F_IV_C_1}
\end{figure}

In this section we discuss the question whether the difference between
strong and weak chaos can be identified from the laser time series
itself. Fig.~\ref{F_IV_C_1} shows two example trajectories (intensity
traces) of a single laser with self-feedback for the regimes of weak
and strong chaos. For both strong and weak chaos, there is a
characteristic structure of high peaks which is significantly higher
for weak chaos than for strong chaos. This may be caused by the larger
coupling strength.

\begin{figure}
  \includegraphics[width=0.4\columnwidth]{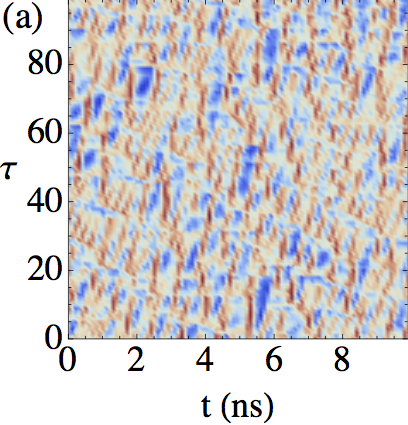}
  \hspace*{0.5cm} \includegraphics[width=0.4\columnwidth]{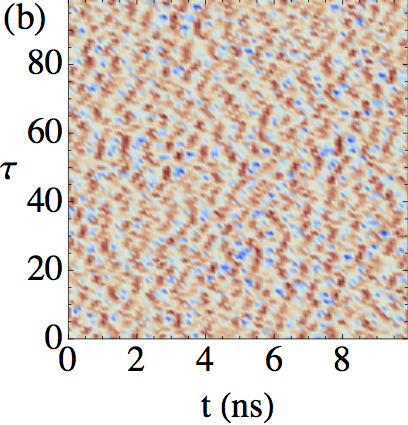}
  \caption{(Color online) Space-time diagram of a laser trajectory
    (reddish: high intensity, blueish: low intensity) for a single
    laser with a self-feedback of $\tau=10\,\mathrm{ns}$ for (a) weak
    chaos ($\sigma = 21 \, \mathrm{ns}^{-1}$) and (b) strong chaos
    ($\sigma = 6 \, \mathrm{ns}^{-1}$) vs the number of the current
    delay window of length $\tau$ (vertical axis) and the time offset
    $t$ in this window (horizontal axis).}
  \label{F_IV_C_4}
\end{figure}

Fig.~\ref{F_IV_C_4} shows the laser trajectories for weak and strong
chaos in space-time diagrams where the vertical axis denotes the
number of the current delay window of length $\tau$ and the horizontal
axis denotes the time offset $t$ in this delay window. In such a
representation, chaos evolves horizontally in space direction for
strong chaos. This is due to the divergence between two nearby
trajectories on the internal time scale of the laser which is much
shorter than the delay time $\tau$. In contrast, weak chaos evolves
vertically in the (discrete) time direction since the divergence
between two nearby trajectories happens on the long time scale of the
delay. Accordingly, the islands of high intensity (red) extend
vertically farther in time direction for weak chaos than they do for
strong chaos. In both kinds of visualization (intensity trace and
space-time diagram), however, one cannot strictly distinguish between
strong and weak chaos in a nonambiguous qualitative way.

\begin{figure}
  \includegraphics[height=0.31\columnwidth]{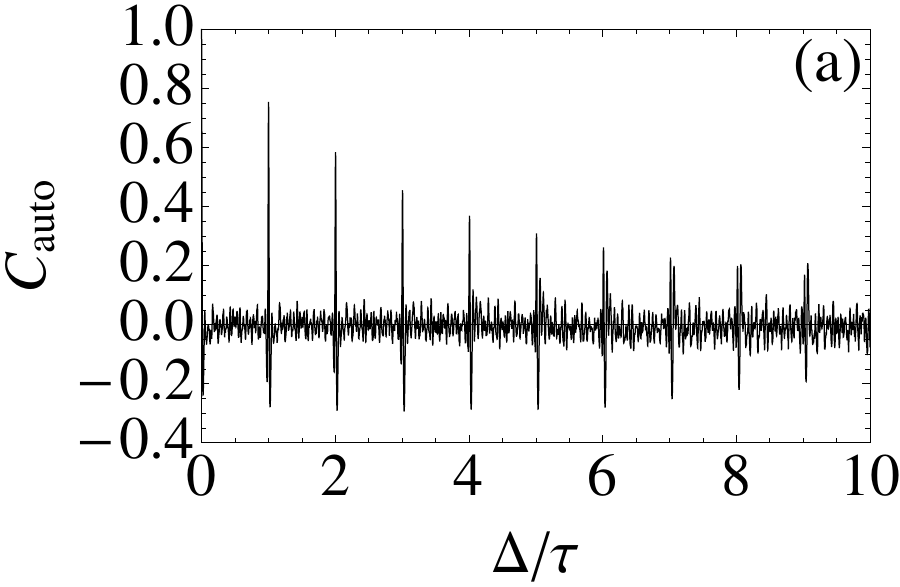}
  \hspace*{\fill} \includegraphics[height=0.31\columnwidth]{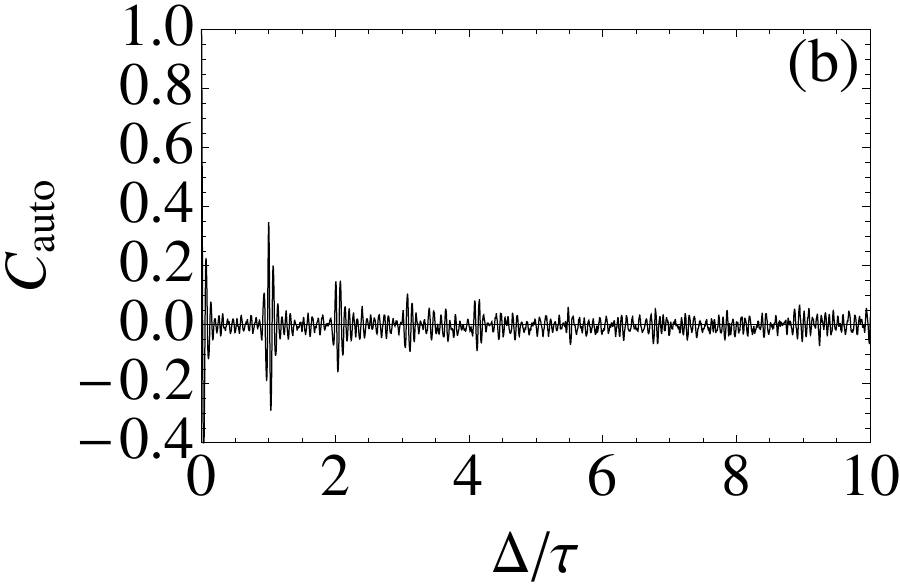}
  \caption{Time-shifted auto-correlations $C_{\mathrm{auto}}$ of a
    single laser with a self-feedback of $\tau=10\,\mathrm{ns}$ for
    (a) weak chaos ($\sigma = 21 \, \mathrm{ns}^{-1}$) and (b) strong
    chaos ($\sigma = 6 \, \mathrm{ns}^{-1}$) vs time shift $\Delta$ in
    units of the delay time $\tau$.}
  \label{F_IV_C_2}
\end{figure}

Fig.~\ref{F_IV_C_2} depicts the time-shifted auto-correlations
$C_{\mathrm{auto}}$ of a single laser with self-feedback for the
regimes of weak and strong chaos. For weak chaos, one clearly sees the
high auto-correlation peaks at multiples of the delay time
$\tau$. Although for strong chaos the chaotic behavior evolves
predominantly on the internal time scale of the laser, whereas for
weak chaos it evolves on the time scale of the delay, there are
non-negligible auto-correlations after multiples of the delay time
$\tau$ even for strong chaos. However, they are significantly smaller
than for weak chaos and decay faster with increasing time shift
$\Delta$.

\begin{figure}
  \includegraphics[height=0.33\columnwidth]{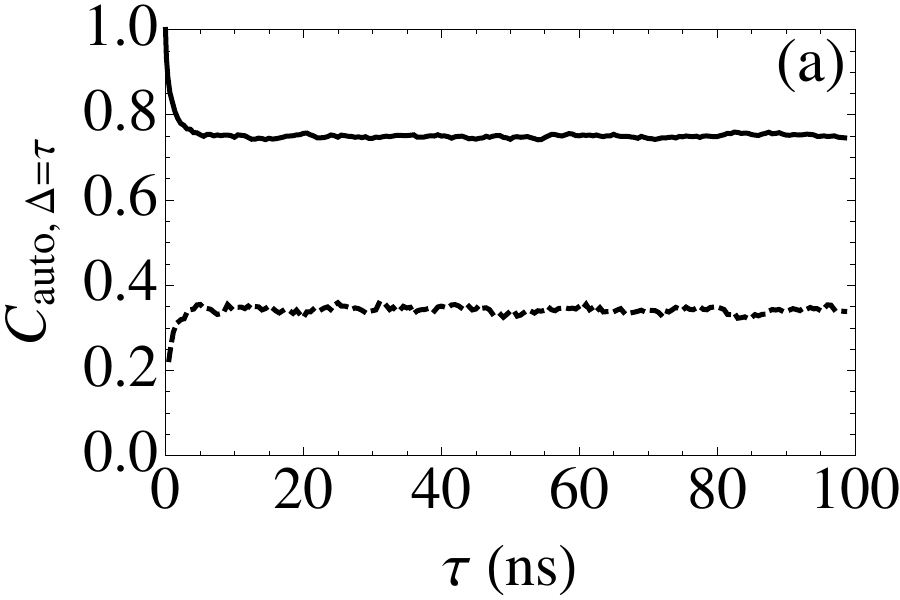}
  \hspace*{\fill} \includegraphics[height=0.33\columnwidth]{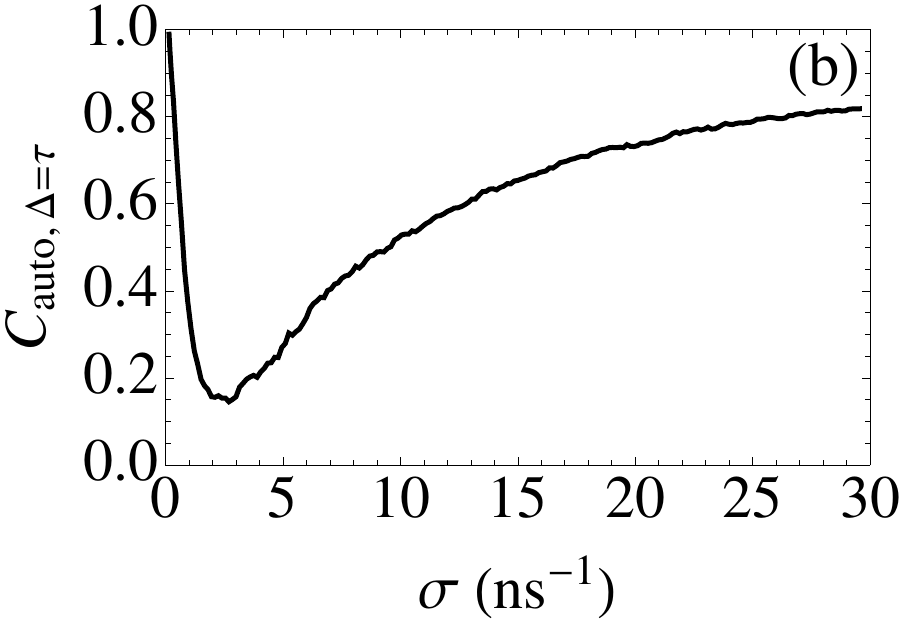}
  \caption{(a)~Auto-correlations $C_{\mathrm{auto}}$ with a time shift
    of $\Delta = \tau$ for a single laser with self-feedback for weak
    chaos ($\sigma = 21 \, \mathrm{ns}^{-1}$, solid line) and strong
    chaos ($\sigma = 6 \, \mathrm{ns}^{-1}$, dashed line) vs delay
    time $\tau$. (b)~Auto-correlations $C_{\mathrm{auto}}$ with a time
    shift of $\Delta = \tau$ for a single laser with a self-feedback
    of $\tau=10\,\mathrm{ns}$ vs coupling strength $\sigma$.}
  \label{F_IV_C_3}
\end{figure}

Considering the auto-correlations $C_{\mathrm{auto},\Delta=\tau}$
after one delay time $\tau$ in Fig.~\ref{F_IV_C_3}(a), we observe that
they do not decay for large delay times with increasing $\tau$ but
remain constant. Surprisingly, this is not only the case for weak
chaos but also for strong chaos. In Fig.~\ref{F_IV_C_3}(b) we depict
the auto-correlations $C_{\mathrm{auto},\Delta=\tau}$ with a time
shift of one delay time $\tau$ in dependence on the coupling strength
$\sigma$. We find no sharp transition at the critical coupling
strengths where the transitions between strong and weak chaos
appears. For weak chaos, however, the auto-correlations
$C_{\mathrm{auto},\Delta=\tau}$ are significantly higher than for
strong chaos.

We conclude that a linear measure like the auto-correlation function
cannot clearly uncover the difference between strong and weak
chaos. Instead, we propose the usage of nonlinear measures in order to
detect a more significant difference in the relationship between
$\mathbf{x}(t)$ and $\mathbf{x}(t-\tau)$. The idea is motivated by the
fact, that in weak chaos for sufficiently large delay times, the
system can be considered to be in a state of generalized
synchronization with its own time-delayed feedback, whereas in strong
chaos the state of the system is formally independent of its
input. Note that this independence does not imply the total absence of
linear correlations, as we have demonstrated. The nature of the
nonlinear functional dependence $\mathbf{x}[\mathbf{x}_{\tau}]$ in
weak chaos is exploited in the corresponding Abarbanel test as
described in Sec.~\ref{subsec:subexp}. Hence, the detection of strong
and weak chaos from time series can be reduced to the problem of
detecting generalized synchronization
\cite{Parlitz2012,Soriano2012}. We refer to the relevant methods used
in this context, like the evaluation of nearest
neighbors~\cite{Koronovskii2011}, mutual information or transfer
entropy~\cite{Schreiber2000}. However, we assume severe computational
difficulties regarding memory usage and runtime connected with the
necessary delay embedding. Additionally, it might even be impossible
to detect generalized synchronization from a finite time series.

\subsection{External cavity modes for strong and weak chaos}

We can relate the strongly or weakly chaotic behavior of a laser
subjected to delayed feedback to the properties of the external cavity
modes (ECMs) of the laser. These ECMs are rotating wave solutions of
the LK equations of the form $E(t)=E_0 \, \mathrm{e}^{\mathrm{i} \,
  \omega \, t}$ and $n(t)=n$ with constant amplitude $E_0$, frequency
$\omega$ and carrier density $n$ of the laser. The spectrum of ECMs is
often represented in the $(\omega,n)$-plane. In this plane the ECMs
lie on an ellipse. ECMs located on the lower half are focus solutions
called modes. The solutions located on the upper half of the ellipse
are saddle points also referred to as antimodes.

Depending on the laser parameters, the ECMs have different stability
properties. In the chaotic regime the ECMs can be seen as the skeleton
of the chaotic attractor \cite{SAN94,MUL98}. In the low frequency
fluctuations regime, which occurs for low pump currents and moderate
to strong coupling strengths, the intensity slowly increases, followed
by a sudden power dropout. During the buildup process, the trajectory
travels along the modes in the direction of the maximal gain mode,
until the trajectory is expelled along the unstable manifold of an
antimode. This causes the power dropout. In the coherence collapse
regime, the dynamics can be described as a chaotic itinerancy between
modes and antimodes.

The stability can be calculated by inserting the ECM solution into
Eq.~\eqref{LKlinearization}. In the long delay limit, the
characteristic equation of a steady state, such as the ECM solutions,
has two types of solutions, which show a different scaling behavior
with the delay time \cite{Lichtner2011}. The strongly unstable
spectrum consists of isolated points, which are approximated by the
unstable eigenvalues of the Jacobian of the LK equations without
delayed terms. These eigenvalues do not scale with the delay. Beside
this strongly unstable spectrum, the characteristic equation has an
infinite number of solutions, forming the pseudocontinuous
spectrum. The real part of these solutions scales inversely with the
delay.

One can thus distinguish between strongly and weakly unstable ECMs in
an analogous way as we distinguish between strong and weak chaos. The
local eigenvalues of the Jacobian without delay terms play a similar
role as the sub-LE $\lambda_0$: Strongly unstable ECMs have unstable
local eigenvalues, and thus a strongly unstable spectrum. The maximal
eigenvalue is approximated by these local eigenvalues and does not
change with the delay. The weakly unstable (and the stable) ECMs have
stable local eigenvalues. The strongly unstable spectrum does not
exist in this case, these ECMs only have a pseudocontinuous spectrum.
Hence, the real part of the maximal eigenvalue scales inversely with
the delay, just like the maximal LE $\lambda_{\mathrm{m}}$ for weak
chaos \cite{Yanchuk2010a}.

\begin{figure}
  \centering
  \includegraphics[width=\columnwidth]{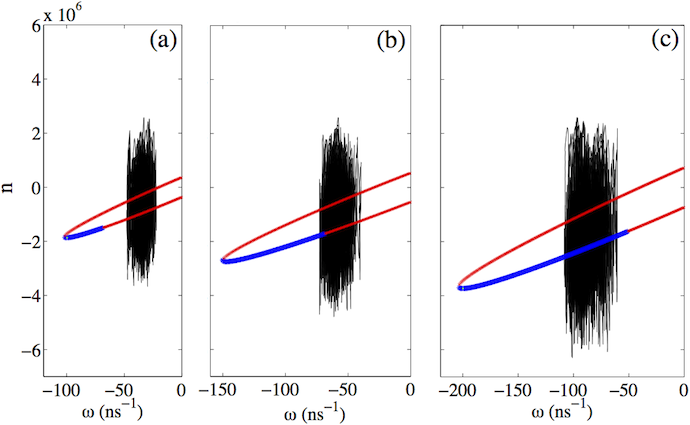}
  \caption{(Color online) Projection of the laser dynamics onto the
    $(\omega,n)$-plane for a delay time $\tau = 10\,\mathrm{ns}$ and a
    pump current $p = 1.10$. Thick blue dots denote weakly unstable or
    stable ECMs, thick red dots represent strongly unstable
    ECMs. Panel (a) shows a strongly chaotic trajectory for $\sigma =
    20 \, \mathrm{ns}^{-1}$. The modes and antimodes involved in the
    dynamics are strongly unstable. In panel (b) the dynamics around
    the transition point between strong and weak chaos is shown for
    $\sigma = 29.28 \, \mathrm{ns}^{-1}$. The trajectory approximates
    the transition point between weakly and strongly unstable modes on
    the ellipse. Panel (c) shows weakly chaotic laser dynamics for
    $\sigma = 40 \, \mathrm{ns}^{-1}$. All the modes involved in the
    dynamics are weakly unstable.}
  \label{fig:ellipsjes}
\end{figure}

In Fig.~\ref{fig:ellipsjes} we show the projection of the laser
dynamics onto the external cavity modes. In the strongly chaotic
regime, all the modes involved in the dynamics are strongly unstable,
as illustrated in Fig.~\ref{fig:ellipsjes}(a). Around the transition
point between strong and weak chaos, a few (two or three) weakly
unstable modes are involved in the dynamics, as shown in
Fig.~\ref{fig:ellipsjes}(b), while most of the modes involved in the
dynamics are weakly unstable in the weakly chaotic regime, as shown in
Fig.~\ref{fig:ellipsjes}(c). These features are independent of the
pump current. The antimodes are always strongly unstable, both in the
weakly and strongly chaotic regime.

\subsection{Sub-Lyapunov exponents for two self-feedbacks with
  time-scale separated delays}

Until now we have only considered a single laser with self-feedback
and have discussed the sub-LE in this context. Now we generalize our
investigation by introducing a second self-feedback which has a delay
time that is much smaller than the first one. The linearized equation
describing the maximal LE is
\begin{equation}
  \dot{\boldsymbol{\delta}\mathbf{x}}=
  \mathrm{D}F(\mathbf{x})\,\delta \mathbf{x} +
  \sigma_{\mathrm{s}} \, \mathrm{D}H(\mathbf{x_{\tau_{\mathrm{s}}}}) \,
  \delta \mathbf{x_{\tau_{\mathrm{s}}}} +
  \sigma_{\mathrm{l}} \,
  \mathrm{D}H(\mathbf{x_{\tau_{\mathrm{l}}}}) \,
  \delta \mathbf{x_{\tau_{\mathrm{l}}}}
\end{equation}
where $\tau_{\mathrm{s}}$ is the delay time of the shorter
self-feedback in comparison with the longer self-feedback with delay
time $\tau_{\mathrm{l}}$. Additionally to the sub-LE $\lambda_0$
defined from Eq.~\eqref{linlambda0} we introduce another sub-LE
$\lambda_{0,\mathrm{s}}$ defined from
\begin{equation}
  \dot{\boldsymbol{\delta}\mathbf{x}}_{\mathbf{0,\mathrm{s}}}=
  \mathrm{D}F(\mathbf{x})\,\delta \mathbf{x_{0,\mathrm{s}}} +
  \sigma_{\mathrm{s}} \, \mathrm{D}H(\mathbf{x_{\tau_{\mathrm{s}}}}) \, \delta \mathbf{x_{0,\mathrm{s},\tau_{\mathrm{s}}}}.
  \label{linlambda0s}
\end{equation}
Thus, for $\lambda_{0,\mathrm{s}}$ we consider a subsystem which
includes the shorter self-feedback but omits the longer one in the
linearized equation. As before, the inserted dynamics $\mathbf{x}(t)$
is the trajectory of the full system which includes both the long and
the short self-feedbacks.

\begin{figure}
  \hspace*{1cm}
  \includegraphics[width=0.55\columnwidth]{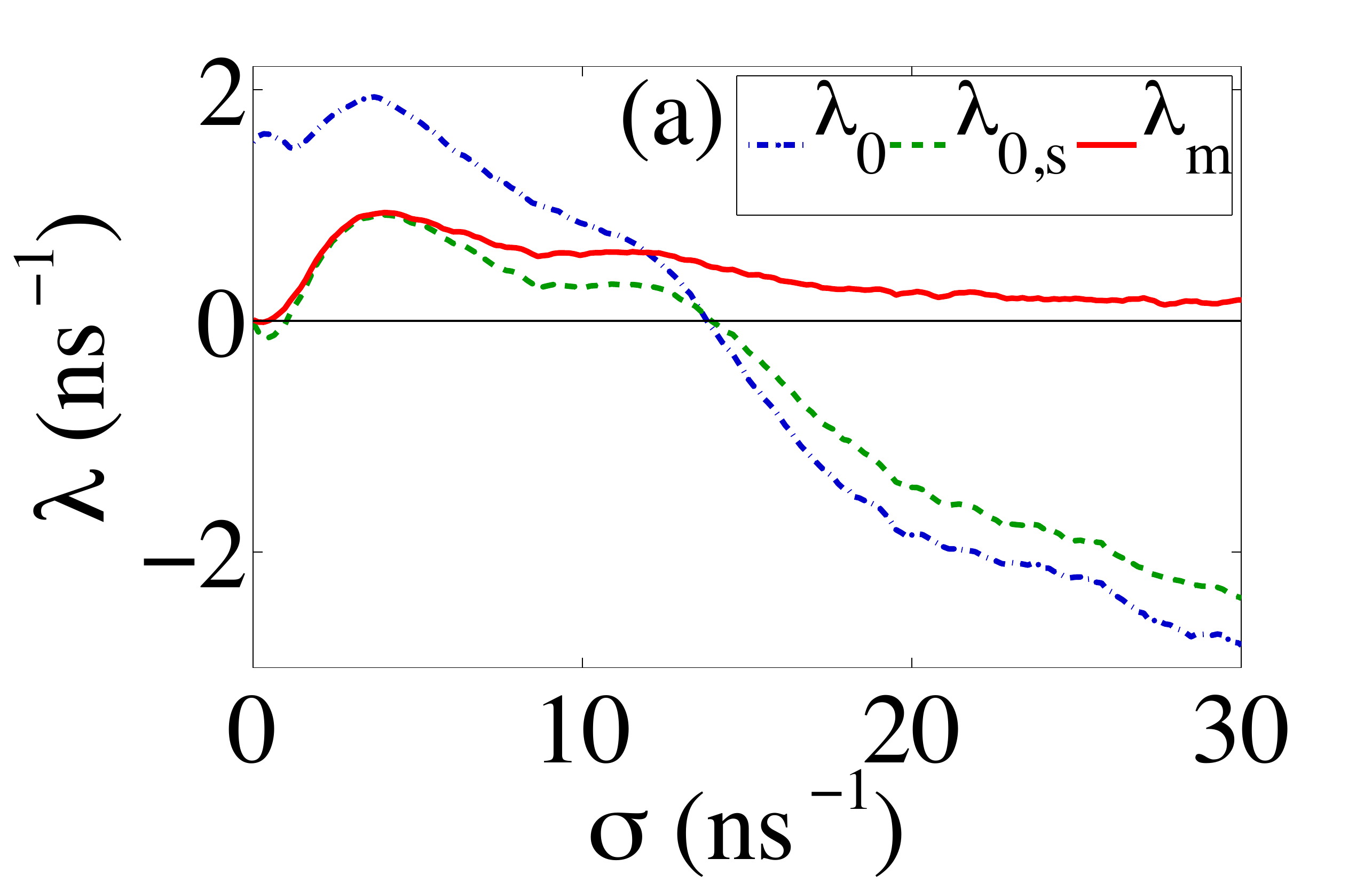}
  \newline
  \includegraphics[width=0.55\columnwidth]{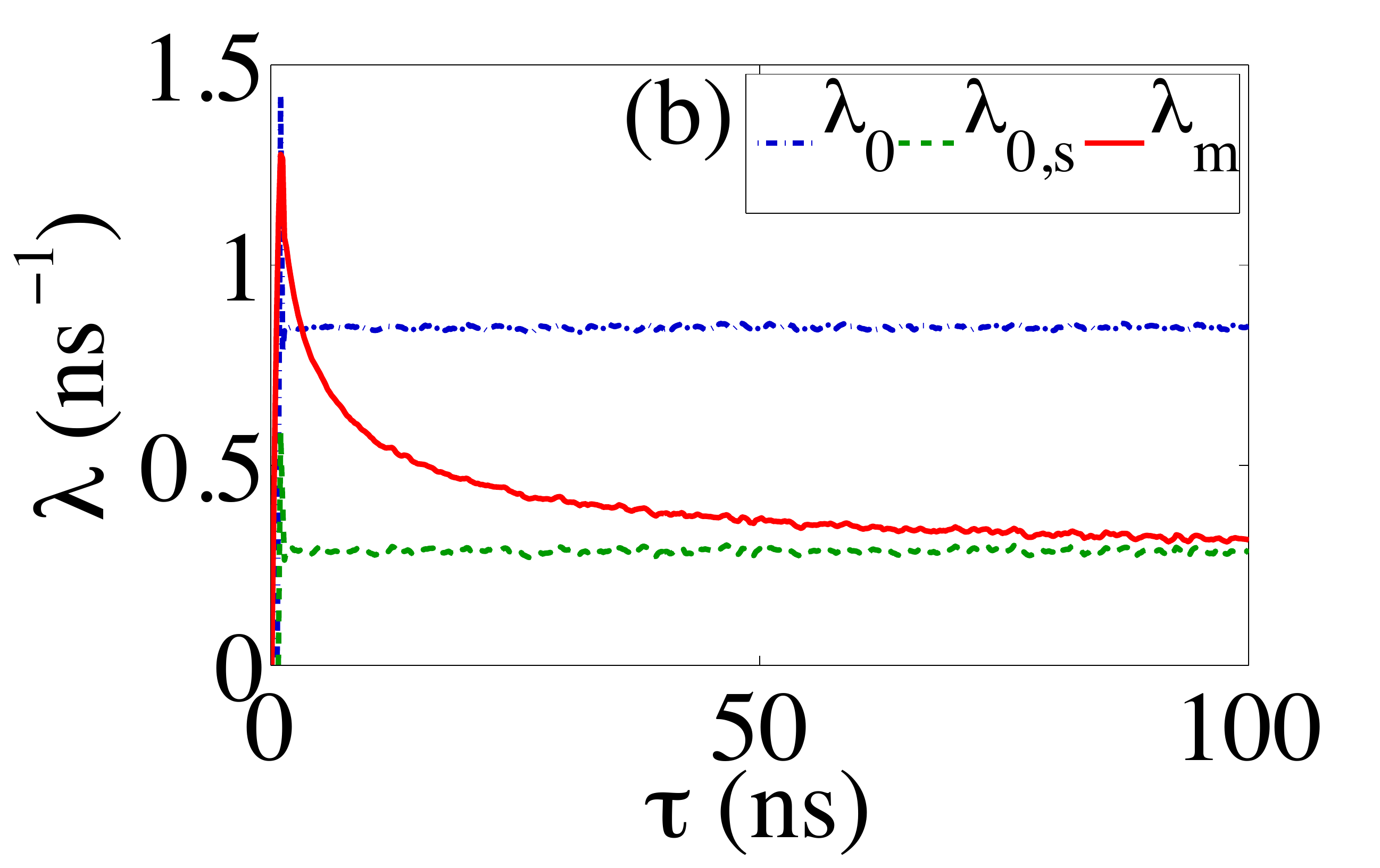}
  \caption{(Color online) Maximal LE $\lambda_{\mathrm{m}}$ (solid
    line) and sub-LEs $\lambda_{0,\mathrm{s}}$ and $\lambda_0$ (dashed
    and dotted lines) for a single laser with two self-feedbacks
    $\tau_{\mathrm{l}}$ and $\tau_{\mathrm{s}}$ for
    (a)~$\tau_{\mathrm{l}} = 10\,\mathrm{ns}$, $\tau_{\mathrm{s}} =
    0.1\,\mathrm{ns}$ and $\sigma_{\mathrm{s}}=5\,\mathrm{ns}^{-1}$ vs
    coupling strength $\sigma:=\sigma_{\mathrm{l}}$ and
    (b)~$\tau_{\mathrm{s}} = 0.1\,\mathrm{ns}$,
    $\sigma_{\mathrm{s}}=5\,\mathrm{ns}^{-1}$ and
    $\sigma_{\mathrm{l}}=10\,\mathrm{ns}^{-1}$ (strong chaos) vs delay
    time $\tau:=\tau_{\mathrm{l}}$.}
  \label{F_IV_E_1}
\end{figure}

Fig.~\ref{F_IV_E_1}(a) shows the sub-LEs $\lambda_0$ and
$\lambda_{0,\mathrm{s}}$ together with the maximal LE
$\lambda_{\mathrm{m}}$ for the case when $\tau_{\mathrm{l}}$ is much
larger than $\tau_{\mathrm{s}}$ such that we have a time scale
separation between the two delays. We observe that $\lambda_{0,s} <
\lambda_0$ holds for strong chaos and $\lambda_{0,s} > \lambda_0$ for
weak chaos. Furthermore, $\sigma_{\mathrm{crit,0}} \approx
\sigma_{\mathrm{crit,0,s}}$ holds for the right transition at large
$\sigma$, i.\,e. both sub-LEs change their signs at approximately the
same coupling strength and can hence both be used as an indicator for
strong or weak chaos there. For small $\sigma$, however, only
$\lambda_{0,\mathrm{s}}$ changes its sign indicating the transitions
between weak and strong chaos there. Thus, $\lambda_{0,s}$ is the new
relevant sub-LE which determines the occurence of strong or weak chaos
in a system with two time-scale separated delays. This is confirmed by
Fig.~\ref{F_IV_E_1}(b) which shows that in the regime of strong chaos,
$\lambda_{\mathrm{m}} \to \lambda_{0,\mathrm{s}}$ holds for increasing
$\tau_{\mathrm{l}}$, while for weak chaos, $\lambda_{\mathrm{m}} \sim
1/\tau_{\mathrm{l}}$.

\section{Networks of lasers with time-delayed couplings}

\subsection{The master stability formalism}

We investigate complete chaotic synchronization of $N$ identical
coupled laser elements. The state of each element is described by
$\mathbf{x}^i$ with $i=1\ldots N$. We start from the physically
motivated ansatz
\begin{equation}
  \dot{\mathbf{x}}^i=\mathbf{F}(\mathbf{x}^i)+\sigma\sum_{j=1}^N G^{ij}\,\mathbf{H}(\mathbf{x}^j_{\tau}).
  \label{lasernetwork}
\end{equation}
The matrix $G\in\mathbb{R}^{N\times N}$ contains the network topology.
$G^{ij}$ is the normalized coupling strength with which the laser $i$
receives input from laser $j$. In order to guarantee the existence of
the completely synchronized state
$\mathbf{x}^1(t)=\mathbf{x}^2(t)=\ldots=\mathbf{x}^N(t)=:\mathbf{s}(t)$,
the row sum of $G$ has to be constant for all rows, and we choose
$\sum_j G^{ij}=1$. Then the synchronized state $\mathbf{s}(t)$ reduces
Eq.~(\ref{lasernetwork}) to
\begin{equation}
  \dot{\mathbf{s}}=\mathbf{F}(\mathbf{s})+\sigma\,\mathbf{H}(\mathbf{s_{\tau}}).
\end{equation}
The stability properties of the synchronized state $\mathbf{s}(t)$ are
well-described by the master stability formalism from Pecora and
Carroll~\cite{Pecora1998}. For completeness, we summarize the main
ideas introduced in their work. A small perturbation is applied to the
synchronized state, such that
$\mathbf{x}^i(t)=\mathbf{s}(t)+\boldsymbol{\delta}\mathbf{x}^i(t)$. The
equations of motion are then linearized around $\mathbf{s}(t)$ and the
perturbations obey the equations of motion
\begin{equation}
  \dot{\boldsymbol{\delta}\mathbf{x}^i}=\mathrm{D}F(\mathbf{s})\,\boldsymbol{\delta}\mathbf{x}^i+\sigma\sum_{j=1}^N G^{ij}\,
  \mathrm{D}H(\mathbf{s}_{\tau})\,\boldsymbol{\delta}\mathbf{x}^i_{\tau}.
  \label{laserlins}
\end{equation}
This set of equations can be decoupled into the amplitudes
$\boldsymbol{\xi}^k(t)$ of the network eigenmodes
$\mathbf{g}^k\in\mathbb{R}^N$, for which
$G\cdot\mathbf{g}^k=\gamma_k\,\mathbf{g}^k$ and $k=1\ldots N$. With
$\boldsymbol{\delta}\mathbf{x}^i=\sum_k g^{k,i}\,\boldsymbol{\xi}^k$
we obtain $N$ decoupled equations
\begin{equation}
  \dot{\boldsymbol{\xi}^k}=\mathrm{D}F(\mathbf{s})\,\boldsymbol{\xi}^k+
  \sigma\,\gamma_k\,\mathrm{D}H(\mathbf{s_{\tau}})\,\boldsymbol{\xi}^k_{\tau}.
  \label{eq_MSF}
\end{equation}
Integration of this equation for the $k$-th perturbation mode
$\mathbf{g}^k$ yields the maximal LE $\lambda_k$, which tells us about
the stability of the mode, meaning that a perturbation
$\boldsymbol{\delta}\mathbf{x}(t)=(\boldsymbol{\delta}\mathbf{x}^1(t),\boldsymbol{\delta}\mathbf{x}^2(t),\ldots,\boldsymbol{\delta}\mathbf{x}^N(t))$
in direction of $\mathbf{g}^k$ grows or decays exponentially. By
construction, there exists at least $\gamma_1=1$ with the eigenvector
$\mathbf{g}^1=(1,1,\ldots,1)^\top$, which corresponds to a
perturbation
$\boldsymbol{\delta}\mathbf{x}(t)=\boldsymbol{\delta}\mathbf{s}(t)$
within the SM. It determines, whether the synchronous dynamics is
chaotic ($\lambda_1>0$) or not. The other modes correspond to linear
combinations of differences between the laser elements and therefore
the necessary condition to find stable complete synchronization is
$\lambda_k<0$ for $k=2\ldots N$. Since all $|\gamma_k|\le 1$, the
knowledge of the master stability function $\lambda(\gamma)$ is
sufficient to predict, whether a network of coupled lasers described
by the coupling matrix $G$ is able to display complete synchronization
or not.

\subsection{Master stability function for weak chaos}

Here we show the general master stability function for the limit of
large delays in weak chaos. To this end, we make use the scaling
behavior $\lambda_{\mathrm{m}}=\mathcal{O}(\tau^{-1})$. The initial
point of our considerations is a generalization of Eq.~(\ref{eq_MSF}),
which reads
\begin{equation}
  \dot{\mathbf{y}}=A(t)\,\mathbf{y}+\kappa\,B(t)\,\mathbf{y_{\tau}},
  \label{eq:yhelpeq}
\end{equation}
where $A(t):=\mathrm{D}F[\mathbf{s}(t)]$ and
$B(t):=\mathrm{D}H[\mathbf{s}(t-\tau)]$. The exponent
$\lambda_{\mathbf{y}}(\kappa)$ provides the master stability
function. Transformation with $\mathbf{z}=\exp(-\lambda_{\mathbf{y}}
\,t)\,\mathbf{y}$ yields a system with exponent
$\lambda_{\mathbf{z}}=0$
\begin{equation}
  \dot{\mathbf{z}}=(A(t)-\lambda_{\mathbf{y}}\cdot\mathbb{1})\,\mathbf{z}+\kappa\,\e^{-\lambda_{\mathbf{y}}\,\tau}\,B(t)\,\mathbf{z_{\tau}}.
  \label{eq:zhelpeq}
\end{equation}
Weak chaos has the important property, that by changing
$A(t)\rightarrow A'(t)=\mathbf{A}(t)+\varepsilon\cdot\mathbb{1}$ with
$\varepsilon\in\mathbb{R}$ and $|\varepsilon|<\lambda_0$, we affect
the LE only in the order of $\tau^{-1}$. Therefore, if $\varepsilon$
itself is decreasing with $\tau$, the effect on the exponent is of a
smaller order in $\tau$, i.\,e. $\mathcal{O}(\tau^p)$ with
$p<-1$. Then for sufficiently large $\tau$, removing
$\lambda_{\mathbf{y}}$ from the first term on the RHS of
Eq.~\eqref{eq:zhelpeq} leads to
\begin{equation}
  \dot{\mathbf{y'}}=A(t)\,\mathbf{y'}+\kappa\,\e^{-\lambda_{\mathbf{y}}\,\tau}\,B(t)\,\mathbf{y'_{\tau}},
  \label{eq:zshelpeq}
\end{equation}
with the exponent
$\lambda'_{\mathbf{y}}=\mathcal{O}(\tau^p)$. Comparison of the
equivalent Eqs.~\eqref{eq:yhelpeq} and \eqref{eq:zshelpeq} shows, that
in the \textit{leading} order of $\tau^{-1}$, rescaling of the
coupling strength
$\kappa\rightarrow\kappa'=\kappa\,\exp(-\lambda_{\mathbf{y}}\,\tau)$
led to an exponent $\lambda'_{\mathbf{y}}\approx 0$. From the
knowledge of the zero-crossing at $\kappa'$, we directly obtain the
scaling law
\begin{equation}
  \lambda_{\mathbf{y}}(\kappa)=\frac{1}{\tau}\ln\frac{\kappa}{\kappa'}.
  \label{eq:lambdascaling}
\end{equation}
Although in general the value of $\kappa'$ is unknown, this
logarithmic law allows us to connect the maximal LE of the SM with the
stability of all transversal modes of a network. We refer to
Eq.~\eqref{eq_MSF}, in which $\kappa=\sigma\,\gamma_k$. Assume the
maximal LE $\lambda_{\mathrm{m}}=\lambda_1$ of the SM is known. It
corresponds to $\gamma_1=1$, so
$\lambda_{\mathrm{m}}=\lambda_1=\lambda(\sigma\cdot1)$ is the point at
which we can fix the master stability function. Then for an arbitrary
$\gamma_k$ we obtain from Eq.~\eqref{eq:lambdascaling} the exponent
\begin{equation}
  \lambda(\sigma\,\gamma_k)=\lambda_{\mathrm{m}}+\ln|\gamma_k|.
\end{equation}
Here we have made use of the fact, that the exponent depends only on
the absolute value $|\kappa|$ for large delays. This scaling relation
also holds for steady states \cite{Flunkert2010}. Transversal
stability of the $k$-th mode is given if
$\lambda(\sigma\,\gamma_k)<0$. This leads to the synchronization
criterion
\begin{equation}
  |\gamma_k|<\e^{-\lambda_{\mathrm{m}}\,\tau},
  \label{eq:synccondition}
\end{equation}
which connects transversal and longitudinal stability in a network
exhibiting weak chaos.

\subsection{Master stability function for the Lang-Kobayashi dynamics}

\begin{figure}
  \includegraphics[width=0.65\columnwidth]{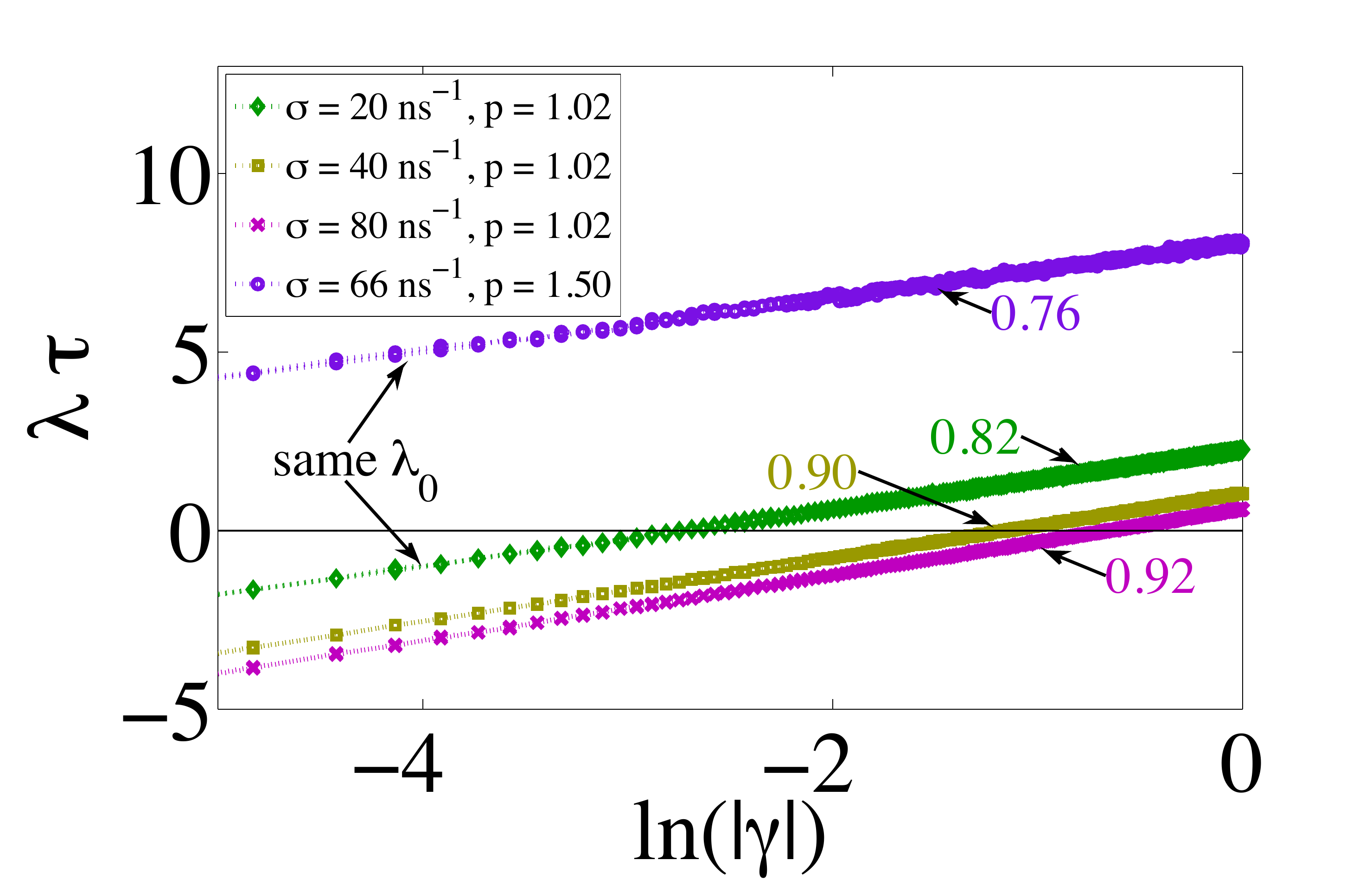}
  \caption{(Color online) Product $\lambda\cdot\tau$ of the master
    stability function $\lambda(\gamma)$ and the delay time
    $\tau=10\,\mathrm{ns}$ for the SM and several different coupling
    strengths $\sigma$ and pump currents $p$ in weak chaos regime vs
    $\ln(|\gamma|)$.}
  \label{F_V_B_1}
\end{figure}

We now investigate the master stability function $\lambda(\gamma)$ for
the LK equations dynamics. For strong chaos, $\lambda(\gamma)$ is
constant since the delay term becomes exponentially small in the
master stability equation \eqref{eq_MSF} for $\tau \to \infty$. Hence,
any perturbation mode with eigenvalue $\gamma$ in a network, or the
mode for complete synchronization in any arbitrary network,
respectively, is unstable for strong chaos if the delay time is
large. Consequently, complete synchronization of networks with
identical, strongly chaotic units is excluded on principle for $\tau
\to \infty$.

For weak chaos, $\lambda(\gamma)$ basically exhibits a logarithmic
dependence on $\gamma$. However, we observe two deviations that can be
considered as effects of the finiteness of the delay time $\tau$ which
is present in every experiment or simulation. As a first deviation,
$\lambda(\gamma)$ does not diverge to $-\infty$ for $\gamma \to 0$ but
has a finite value $\lambda(0)$: the sub-LE $\lambda_0$. It is
important to note here that the sub-LE $\lambda_0$ is equal to the LE
of any perturbation mode with $\gamma=0$ of a network (e.\,g. the mode
of complete synchronization of the two outer lasers in a
bidirectionally coupled chain of three lasers) since then the delay
term drops out in the master stability equation \eqref{eq_MSF},
too. This is also the case in the experimental setup for the
measurement of the sub-LE presented in Fig.~\ref{F_III_A_1}(a). There,
the mode of complete synchronization has $\gamma = 0$, and hence
synchronization between the two laser happens if and only if they show
weakly chaotic behavior.

As a second deviation for weak chaos, Fig.~\ref{F_V_B_1} shows for
several different coupling strengths and pump currents that the
exponent $\nu$ in $\lambda \sim \ln(|\gamma|^{\nu})=\nu \cdot
\ln(|\gamma|)$ is not exactly One as for simple analytic models with
constant coefficients, but rather shows some deviation to exponents
smaller than One. We were able to find this deviation also for the
Lorenz dynamics. For constant delay time $\tau$ and constant laser
pump current $p$, the exponent gets closer to One the larger $\sigma$
is, i.\,e. the weaker the chaos becomes. Additionally, if the coupling
strength $\sigma$ is adjusted such that the sub-LE $\lambda_0$ is the
same for two different pump currents, then the slope is closer to One
for a small pump current than for a higher pump current. As presented
in the previous section, a condition for the prediction of
synchronization in a network, Eq.~\eqref{eq:synccondition}, can be
derived from the behavior of $\lambda(\gamma)$. Due to the empirically
observed slope smaller than One, this condition should be refined to

\begin{equation}
  |\gamma_2|<\mathrm{e}^{-\lambda_{\mathrm{m}} \, \tau / \nu}
\end{equation}

where $\gamma_2$ is the eigenvalue of the coupling matrix $G$ with the
second largest modulus. This equation allows for a more precise
practical prediction of synchronization in networks of semiconductor
lasers, derived from LK equation modeling.

\subsection{Experimental evidence of strong and weak chaos in
  bidirectionally delay-coupled lasers}
\label{sec:experiment}

In a system with two bidirectionally delay-coupled lasers without
self-feedback or multiple delays, identical chaos synchronization is
not stable due to symmetry breaking \cite{Heil2001}. However, chaos
synchronization can still exist in the generalized sense
\cite{Abarbanel1996}. Here, we give experimental and numerical
evidence that the implications of strong and weak chaos regimes also
apply to the case of generalized chaos synchronization.

\begin{figure}
  \includegraphics[width=0.75\columnwidth]{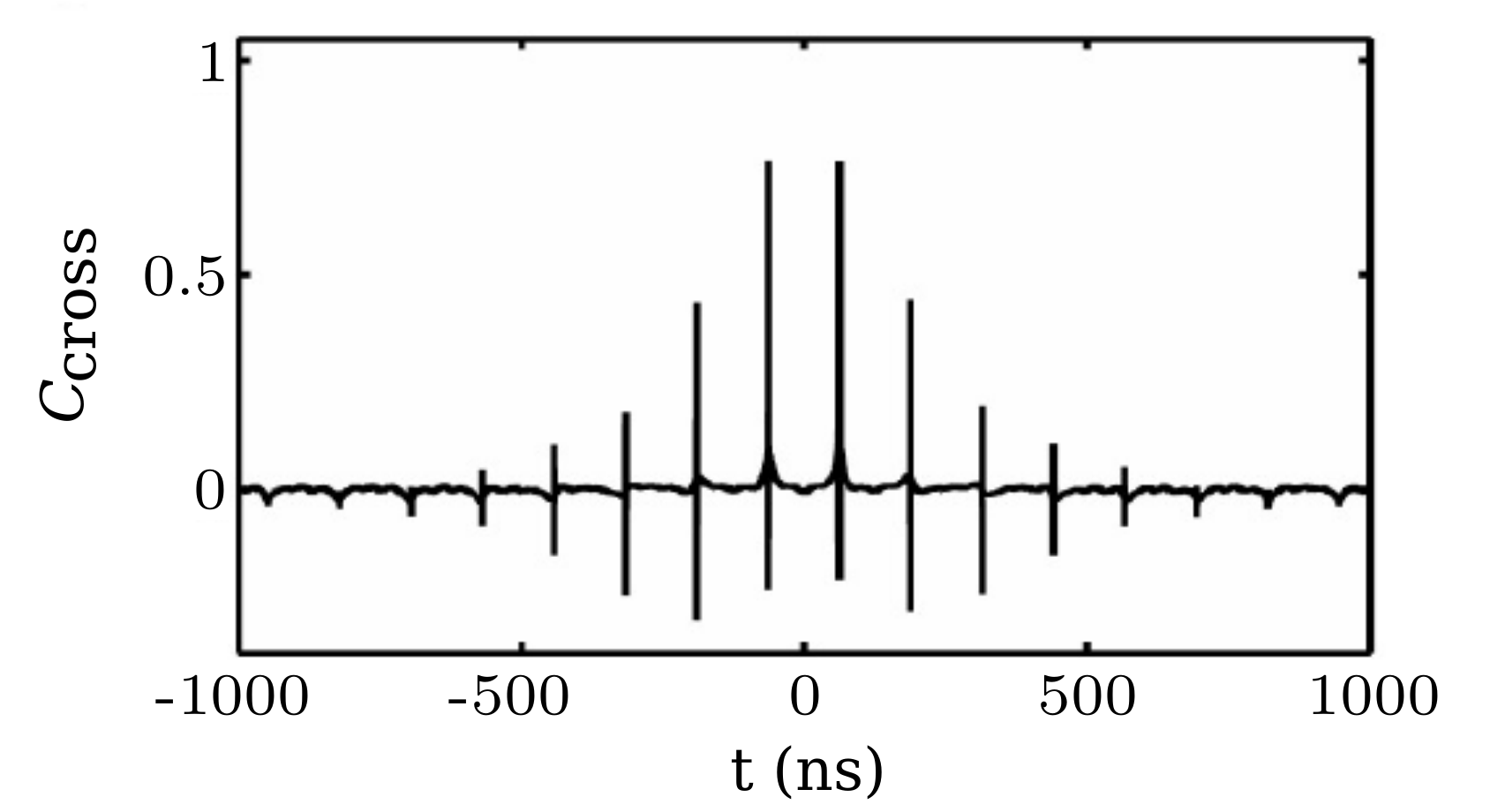}
  \caption{Cross-correlation function of two bidirectionally
    delay-coupled lasers in the chaotic regime
    ($\tau=63\,\mathrm{ns}$).}
  \label{F_V_C_3}
\end{figure}

In the context of delay-coupled elements, correlation measures can
fail to detect synchronization if the number of coupled elements is
large \cite{Soriano2012}. However, the cross-correlation function is
still a good indicator to identify generalized synchronization between
two delay-coupled lasers. We present an example of such a
cross-correlation function in Fig.~\ref{F_V_C_3}, which shows distinct
peaks at the delay time $\tau$ and its odd multiples. In this figure,
the large correlation peak at the delay time indicates that the lasers
are generally synchronized. We argue that generalized chaos
synchronization is only possible, if the lasers are operating in the
weak chaos regime. In contrast, a low correlation peak at a time-lag
$\tau$ is to be expected, if the lasers are operating in the strong
chaos regime.

\begin{figure}
  \includegraphics[width=0.95\columnwidth]{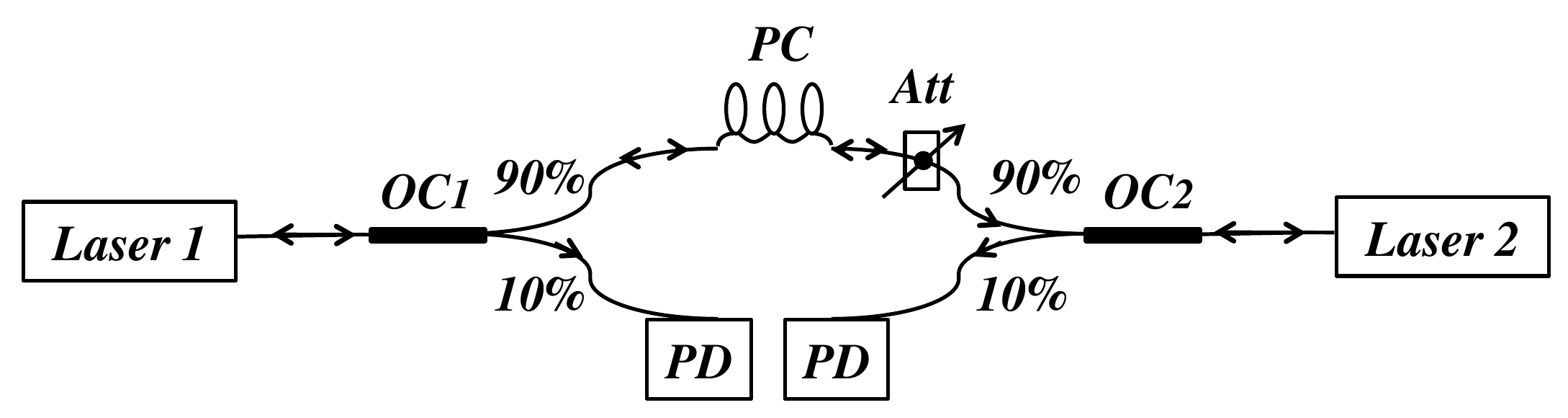}
  \caption{Fiber-based experimental setup of two mutually
    delay-coupled semiconductor lasers. $OC_{1,2}$ stand for the
    optical couplers, $PC$ is the polarization controller, $Att$ is
    the variable attenuator, and $PD$ denotes the photodetectors.}
  \label{F_V_C_2}
\end{figure}

Our fiber-optics-based experimental arrangement is schematically shown
in Fig.~\ref{F_V_C_2}.  We use two single-mode, fiber-pigtailed,
discrete-mode semiconductor lasers, emitting at
$1542\,\mathrm{nm}$. The lasers have been hand-selected in order to
achieve well-matched parameters. The laser temperatures and currents
are stabilized to an accuracy of $0.01\,\mathrm{K}$ and
$0.01\,\mathrm{mA}$, respectively. The lasers are biased at a current
of $1.25\,I_{\mathrm{th}}$, with $I_{\mathrm{th}} = 11.7\,\mathrm{mA}$
being the solitary laser threshold current. As shown in
Fig.~\ref{F_V_C_2}, the coupling path includes two 90/10 optical
couplers $OC_{1,2}$, a polarization controller $PC$, and an optical
attenuator $Att$. The maximum mutual coupling obtained in this
experimental arrangement can be estimated to be $\sim 40\%$ of the
emitted light. The $10\%$ outputs of $OC_{1,2}$ are used for
detection.

\begin{figure}
  \includegraphics[width=\columnwidth]{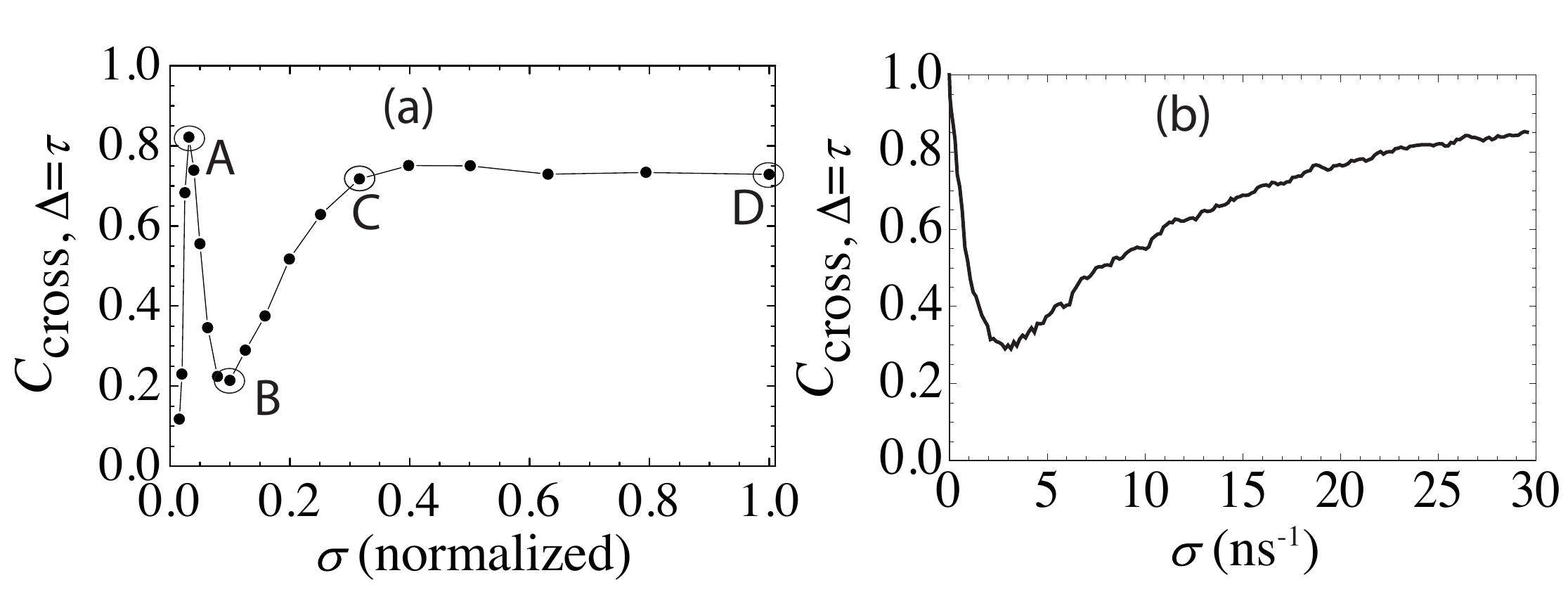}
  \caption{(a) Experimentally measured and (b) numerically calculated
    cross-correlations $C_{\mathrm{cross}}$ with a time shift of
    $\Delta = \tau$ for a bidirectionally coupled pair of lasers as a
    function of the coupling strength $\sigma$. The delay time is
    $\tau=10\,\mathrm{ns}$ in the numerics and $\tau=63\,\mathrm{ns}$
    in the experiments, corresponding to the long delay limit. The
    experimental coupling strength is normalized to the maximum
    coupling obtained in the setup, which is about $40\%$ of the
    emitted light.}
  \label{F_V_C_1}
\end{figure}

The values of the cross-correlation between the intensities emitted by
Lasers 1 and 2 for a time shift of $\tau$ are shown in
Fig.~\ref{F_V_C_1}(a) as a function of the coupling strength
$\sigma$. For strong couplings $\sigma>0.3$, a region of large
correlation is found. Decreasing the coupling strength to
$0.1<\sigma<0.3$ results in a sudden decrease of the correlation. A
second region of large correlation can be identified for weak
couplings $\sigma\sim 0.05$. The correlation decreases again for the
weakest couplings $\sigma<0.05$. The numerical results, shown in
Fig.~\ref{F_V_C_1}(b), agree with the experimental results. For large
couplings, the two coupled lasers are highly correlated. A distinct
region of low correlation can be seen for intermediate couplings,
while the correlation increases again for the weakest coupling
strengths. The numerical simulations do not reproduce the correlation
decrease towards zero coupling since spontaneous emission noise
sources are not considered. The numerical results of the time-shifted
cross-correlations for the two bidirectionally coupled lasers shown in
Fig.~\ref{F_V_C_1}(b) are similar to the auto-correlations of a single
laser with self-feedback shown in Fig.~\ref{F_IV_C_3}(b). The
parameters used in the numerical simulations are listed in the
appendix.

\begin{figure}
  \includegraphics[width=0.95\columnwidth]{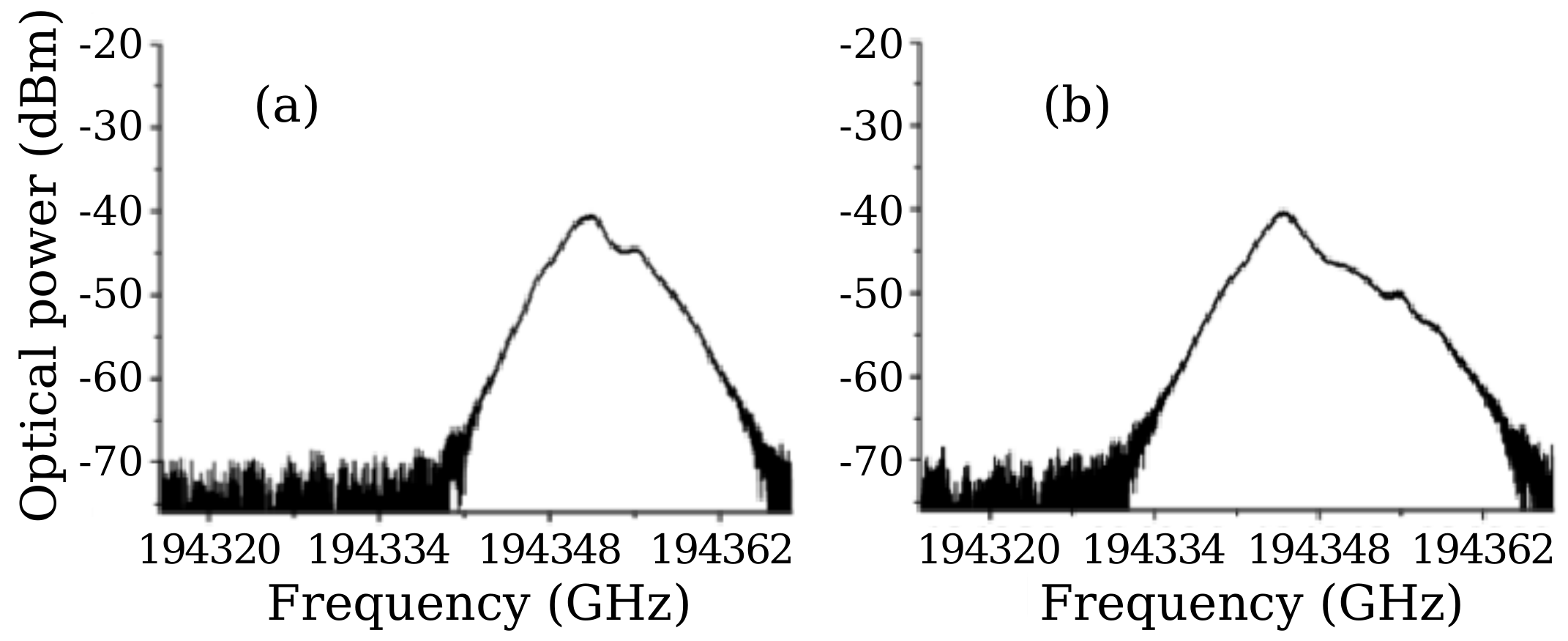}
  \caption{Optical spectra for points denoted as (a) $B$ and (b) $C$
    in Fig.~\ref{F_V_C_1}(a).}
  \label{F_V_C_4}
\end{figure}

Several dynamical states are being observed for variation of the
coupling along the correlation curve in Fig.~\ref{F_V_C_1}(a). The
lasers operate in continuous-wave with noisy fluctuations in the
absence of coupling. For an increasing coupling strength, the lasers
follow a quasi-periodic route via undamped relaxation oscillations
reaching a chaotic state at point $A$. The delay-induced dynamics
produces a dramatic increase in the laser optical linewidth, which
increases from a few MHz to several GHz due to the coupling. We
present in Fig.~\ref{F_V_C_4} the typical shape of the optical spectra
of the chaotic laser for points denoted as $B$ and $C$ in
Fig.~\ref{F_V_C_1}(a). These two spectra are qualitatively similar
apart from their different width, but correspond to significantly
different correlation and synchronization properties. In order to
characterize the optical spectra, we have measured their width,
defined as the $-20\,\mathrm{dB}$ frequency width. The
$-20\,\mathrm{dB}$ width of the optical spectrum is $9\,\mathrm{GHz}$,
$20\,\mathrm{GHz}$, $26\,\mathrm{GHz}$, and $35\,\mathrm{GHz}$, for
points $A$, $B$, $C$, and $D$, respectively.

We observe irregular (chaotic) dynamics in the whole range from $A$ to
$D$.  However, the correlation plot indicates qualitative transitions
of the synchronization behavior between $A$ and $B$, and $B$ and $C$,
respectively.  More precisely, the experimental correlation plot shows
a large correlation around $A$, and from $C$ to $D$, indicating that
the lasers are generally synchronized in these regions. The degree of
synchronizability stems from the dynamical regime the lasers are
operating in. Therefore, we can infer that the lasers operate in a
weak chaos regime in the two high correlation (synchronization)
regions, i.\,e. for $\sigma > 0.3$ and $\sigma \sim 0.05$. In
contrast, a region of low correlation measure can be seen for
intermediate couplings $\sigma \sim 0.1$. This can be associated with
a strong chaos dynamical regime. The numerical correlation plot shows
a similar clear distinction between different correlation regions,
with a window of low correlation around $\sigma \sim
2.5\,\mathrm{ns^{-1}}$. This low correlation can be linked to a strong
chaos dynamical regime.

Our experimental results on two delay-coupled semiconductor lasers
presented here support the sequence `weak to strong to weak chaos'
with an increasing coupling strength. Even though the master stability
function cannot be directly applied to the generally synchronized
solution, the influence of the dynamical regime on the synchronization
is clearly substantiated by our experimental and numerical results.

\subsection{Networks with several distinct sub-Lyapunov exponents}

\begin{figure}
  \begin{minipage}{0.44\columnwidth}
    \includegraphics[width=\columnwidth]{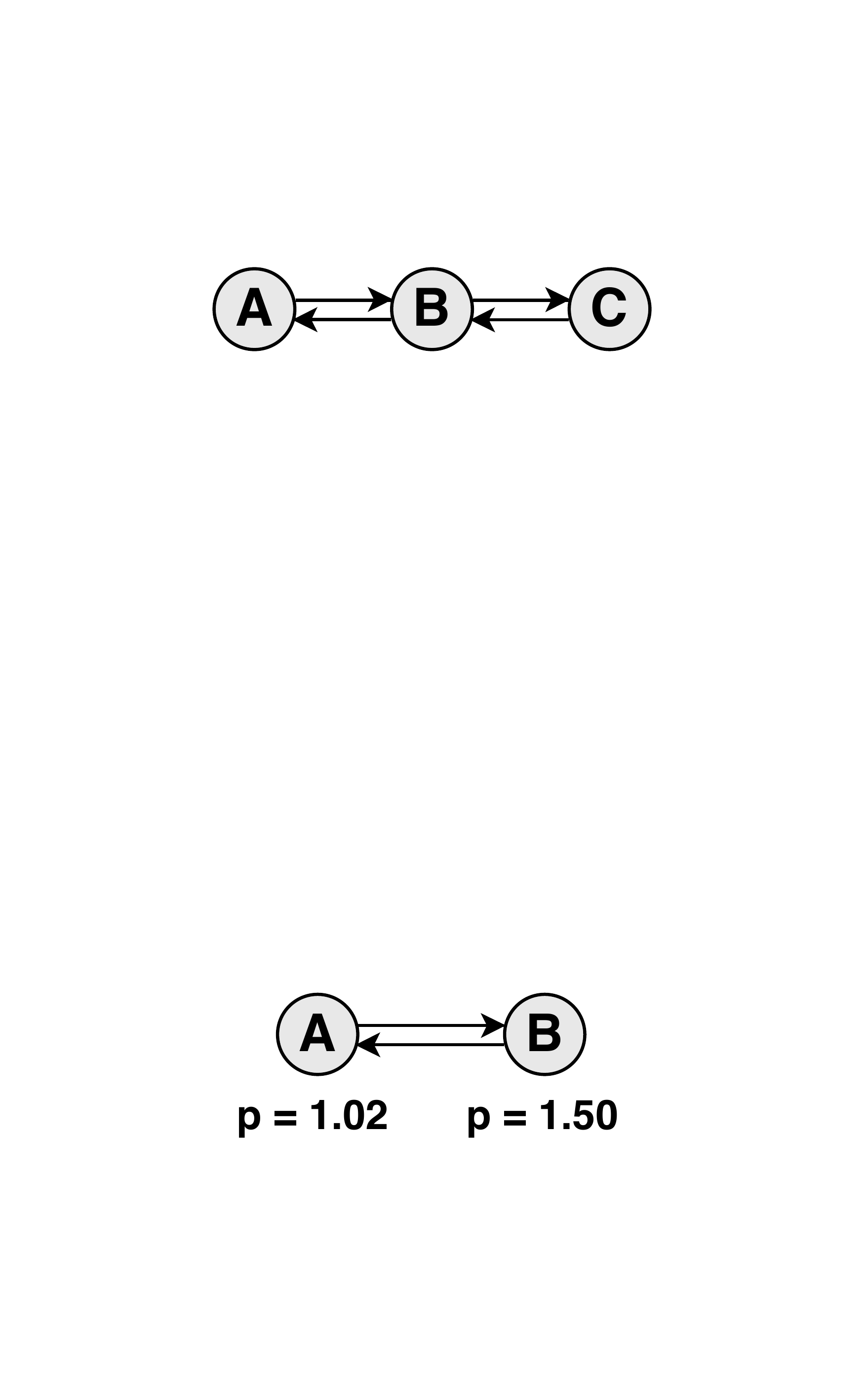}
  \end{minipage}
  \begin{minipage}{0.54\columnwidth}
    \hspace*{\fill}
    \includegraphics[width=0.96\columnwidth]{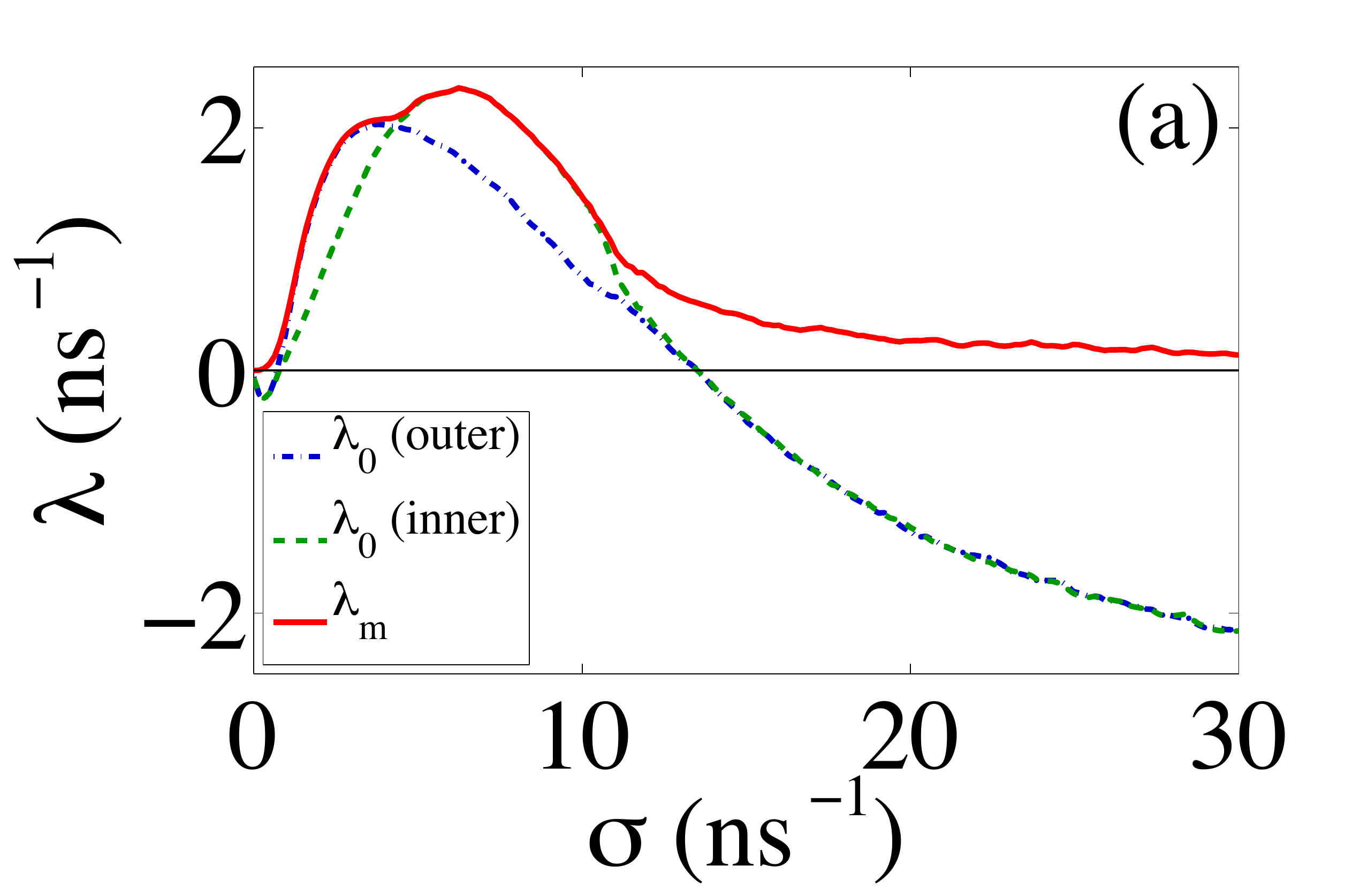}
    \newline
    \includegraphics[width=\columnwidth]{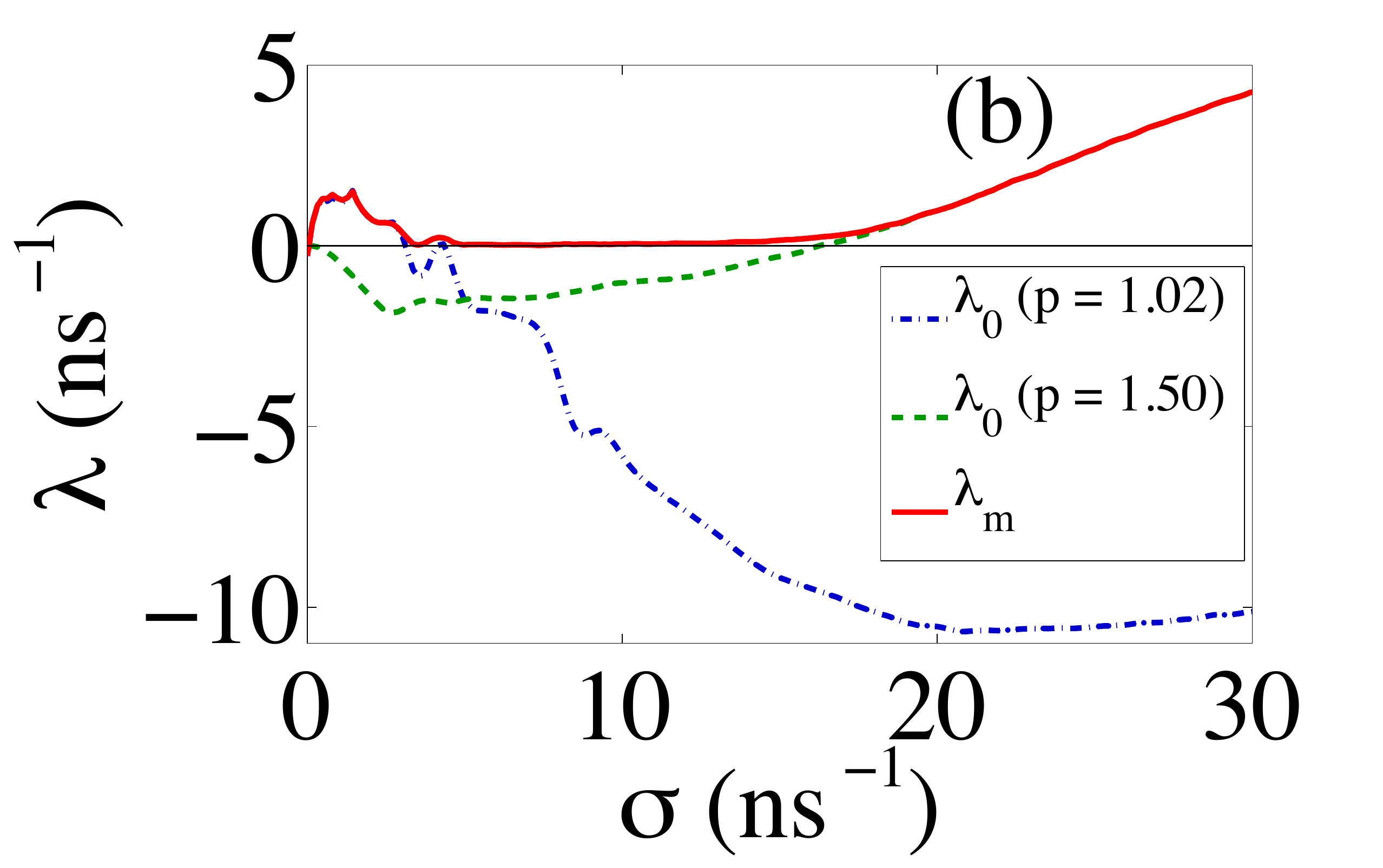}
  \end{minipage}
  \caption{(Color online) (a)~Maximal LE $\lambda_{\mathrm{m}}$ (solid
    line) and sub-LEs $\lambda_{0}$ of the inner laser (dashed line)
    and the outer lasers (dot-dashed line) of a chain of three lasers
    (see left) for a coupling delay time of $\tau=10\,\mathrm{ns}$ vs
    coupling strength $\sigma$.  (b)~Maximal LE $\lambda_{\mathrm{m}}$
    (solid line) and sub-LEs $\lambda_{0}$ (dashed and dot-dashed
    lines) of pair of two lasers with distinct pump currents (see
    left) for a coupling delay time of $\tau=10\,\mathrm{ns}$ vs
    coupling strength $\sigma$.}
  \label{F_V_D_1}
\end{figure}

In a network we can define a sub-LE for each individual unit. If the
network is not completely symmetric then these sub-LEs may differ from
each other, even if the units by themselves are identical. For
example, in a chain of three bidirectionally coupled lasers, the
middle laser is in a different coupling situation than the outer
lasers. It receives input from two lasers while the outer ones receive
input only from one laser. It is important to note here that the
accumulated coupling strength of the inputs, however, is constant for
each laser of the chain. The occurence of strong or weak chaos now
depends on both sub-LEs present in the network. Fig.~\ref{F_V_D_1}(a)
shows that this dependence is simple: The maximal sub-LE of the
network determines whether there is strongly or weakly chaotic
behavior of the complete network's maximal LE. If the largest
$\lambda_0$ in the network is negative then there is weak chaos, if
the largest $\lambda_0$ in the network is positive, then the maximal
LE $\lambda_{\mathrm{m}}$ of the network converges to the maximal
sub-LE of the network. This means that potentially additionally
present strongly chaotic units do not further increase the value to
which the complete network's maximal LE converges for strong chaos.

Additionally to the case when the network is not completely symmetric,
also the individual units in the network may be nonidentical,
e.\,g. in a pair of two bidirectionally coupled lasers with different
pump currents, Fig.~\ref{F_V_D_1}(b). One laser has a pump current
$p=1.02$ and the other has $p=1.50$. It is confirmed that even for
this case, the largest sub-LE of the network determines the maximal LE
of the network and that for strong chaos the maximal LE of the network
follows the largest sub-LE. We further observe that, by making the
lasers in the network nonidentical, we can even induce more
transitions between strong and weak chaos than the transition
weak/strong/weak chaos for a single laser or a network of identical
lasers as shown before. These additional transitions are interrupted
by intervals of $\sigma$ where the maximal LE $\lambda_{\mathrm{m}}$
decays to Zero, i.\,e. windows of periodic behavior can be found.

As a consequence, we conclude that complete synchronization is
excluded on principle for arbitrary networks which contain at least
one strongly chaotic unit if the delay time is very large. Cluster
synchronization, however, is still possible. In this case, the weakly
chaotic units of the network may form one or more clusters. These
clusters are driven by the strongly chaotic units of the network.

\subsection{Sub-Lyapunov exponents for certain network patterns}

In the previous section we have shown that different number of inputs
from other lasers and their arrangement can have the effect of
distinct sub-LEs. This is the case even if the lasers by themselves
are identical. In this section we address the question if the number
of inputs is the only criterion which determines the sub-LE for
otherwise constant system parameters. We would like to emphasize once
more that the accumulated coupling strength of the inputs is constant
for all considered network patterns.

\begin{figure}
  \begin{minipage}{0.44\columnwidth}
    \includegraphics[width=\columnwidth]{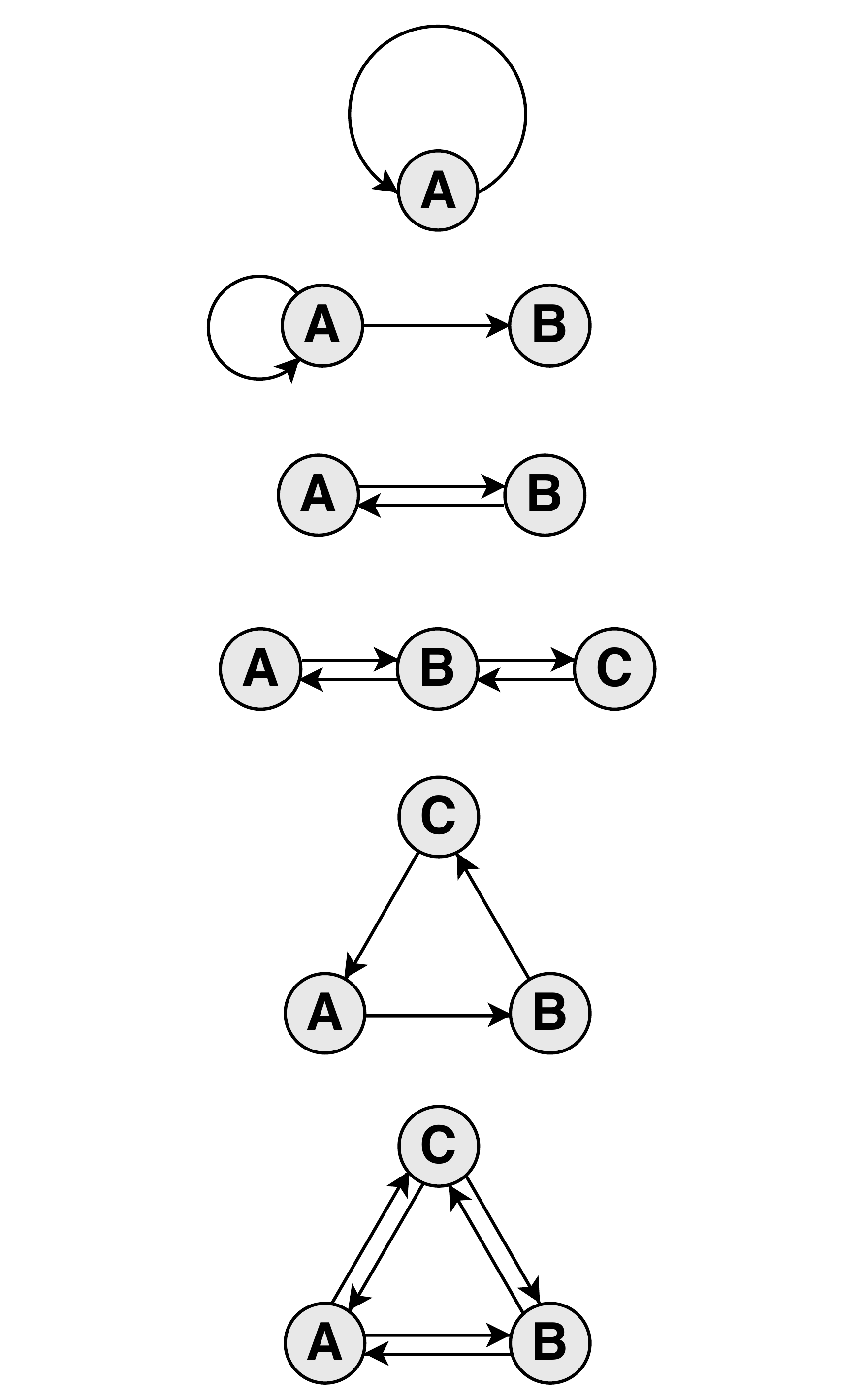}
  \end{minipage}
  \begin{minipage}{0.54\columnwidth}
    \includegraphics[width=\columnwidth]{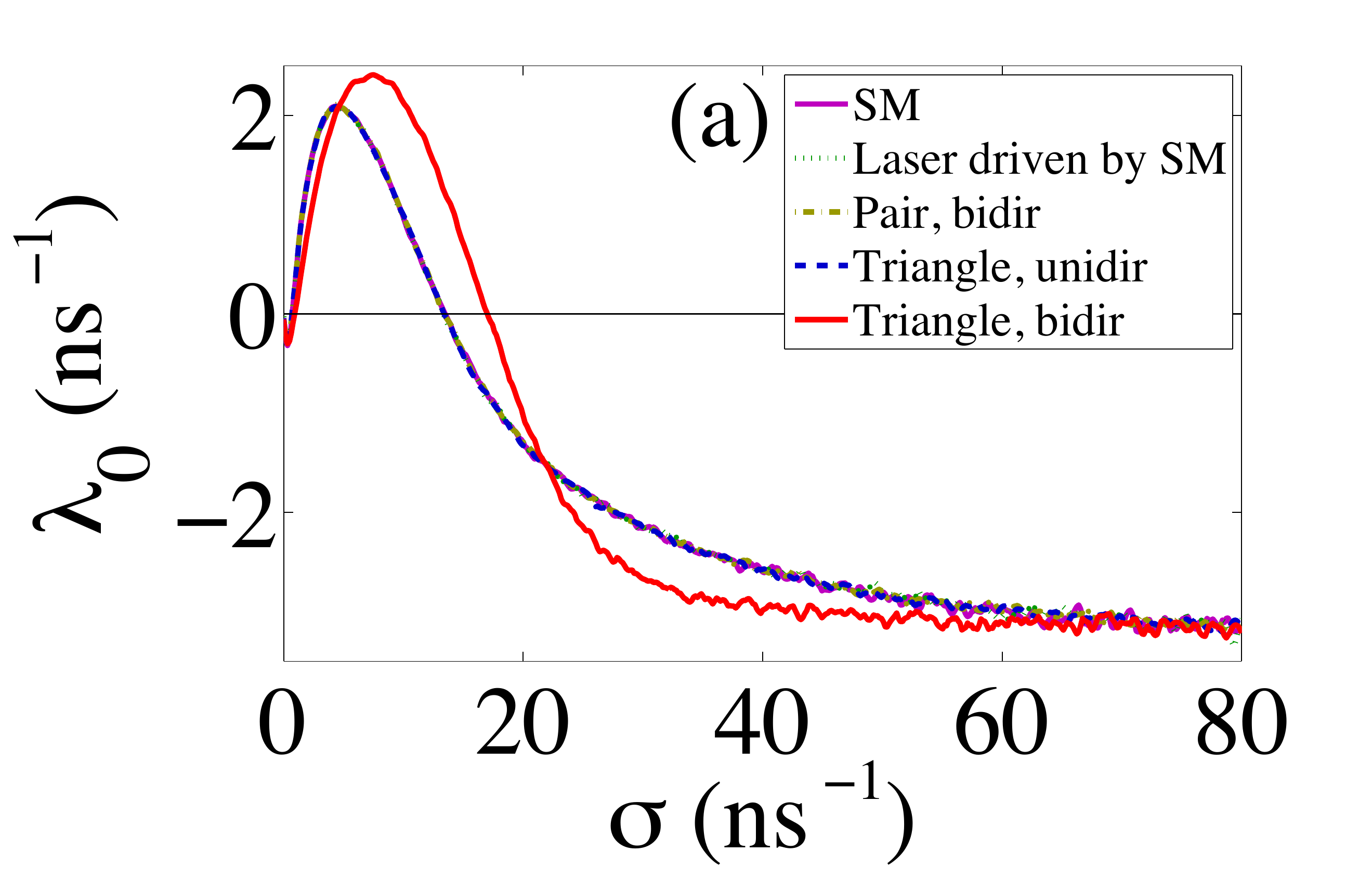}
    \newline
    \includegraphics[width=\columnwidth]{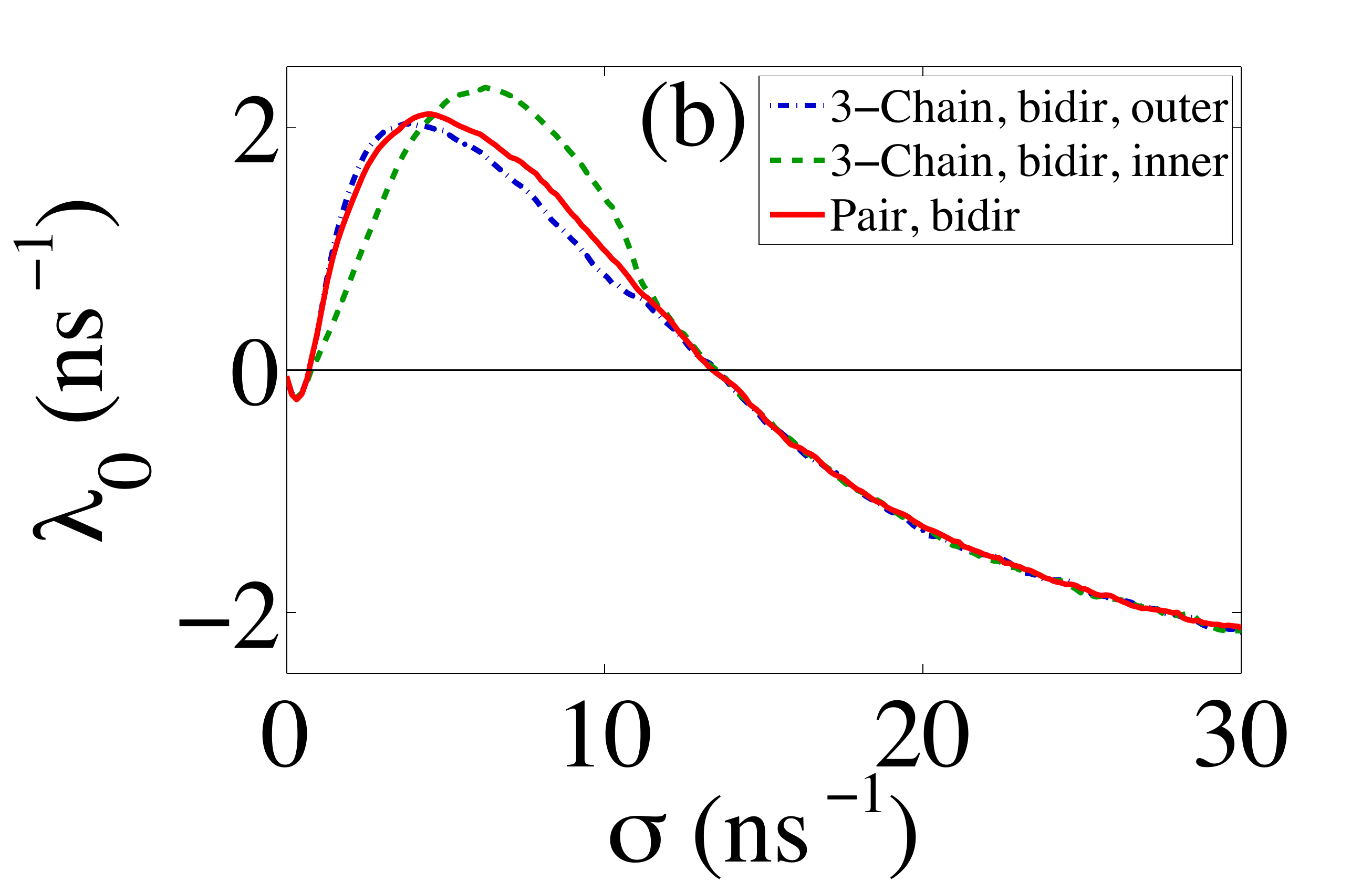}
    \newline \hspace*{\fill}
    \includegraphics[width=0.96\columnwidth]{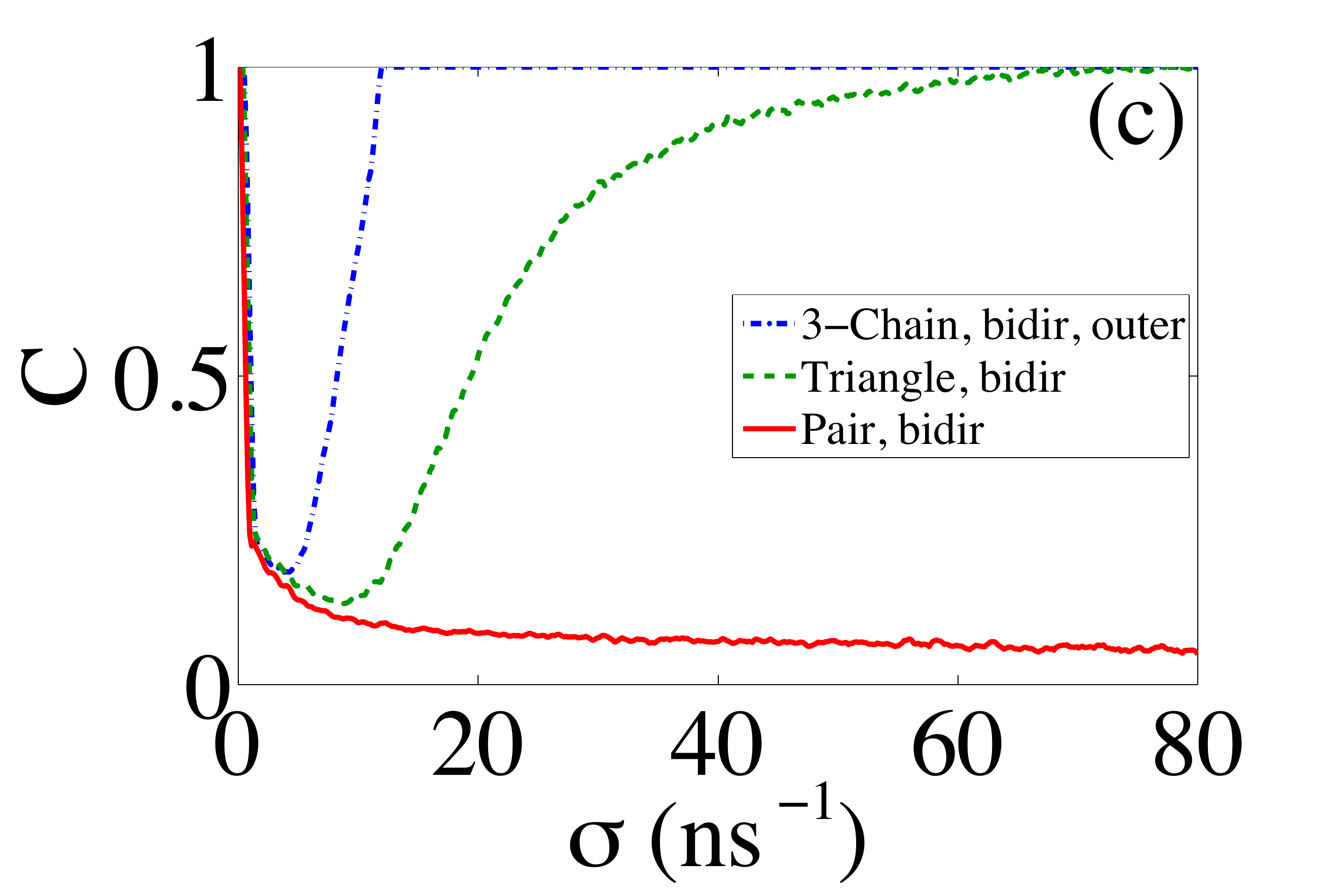}
  \end{minipage}
  \caption{(Color online) (a)~Sub-LEs $\lambda_{0}$ for several
    network patterns where the lasers receive input from one laser
    (see left) in comparison with $\lambda_{0}$ of a bidirectional
    triangle of lasers for a coupling delay time
    $\tau=10\,\mathrm{ns}$ vs coupling strength $\sigma$.  (b)~Sub-LEs
    $\lambda_{0}$ of the inner and outer lasers of a chain of three
    lasers in comparison with $\lambda_{0}$ of a pair of lasers (see
    left) for a coupling delay time $\tau=10\,\mathrm{ns}$ vs coupling
    strength $\sigma$. (c)~Cross-correlation $C$ between the lasers in
    several network patterns for a coupling delay time
    $\tau=10\,\mathrm{ns}$ vs coupling strength $\sigma$.}
  \label{F_V_E_1}
\end{figure}

Fig.~\ref{F_V_E_1}(a) shows a comparison of the sub-LEs of several
networks where each laser receives input from exactly one laser. The
SM, represented by a single laser with self-feedback, and the
bidirectionally coupled pair, for which the experimental evidence of
strong and weak chaos has been provided in Sec.~\ref{sec:experiment},
can be seen as limit cases of unidirectional rings containing one and
two lasers. They are compared with a unidirectional triangle which is
a unidirectional ring of three lasers. The bidirectional pair and the
unidirectional triangle have in common that they do not have an
eigenvalue gap. Hence, they cannot synchronize isochronically for any
coupling strength $\sigma$ as shown in Fig.~\ref{F_V_E_1}(c) for the
example of the bidirectional pair. Nevertheless, their sub-LEs are
exactly the same as the one of the SM represented by a single laser
with self-feedback. Also the sub-LE of Laser $\mathcal{B}$ of the
configuration presented in Fig.~\ref{F_III_A_1}(a) is equal to the
ones mentioned before even in the strong chaos regime where the Laser
$\mathcal{B}$ is not synchronized with Laser $\mathcal{A}$. The reason
for this is that in all cases the lasers receive coherent input from a
single source which exhibits the specific statistics of a laser
trajectory. It does not matter if the statistics comes from a
synchronized laser or an unsynchronized one, as long as it is typical
for a single laser. This can also be nicely seen in
Fig.~\ref{F_V_E_1}(a) by the sub-LE of a bidirectionally coupled
triangle. If the coupling strength $\sigma$ is not large enough to
induce synchronization of this triangle configuration, as seen in
Fig.~\ref{F_V_E_1}(c), then each laser receives incoherently
superimposed input from two unsynchronized lasers. Hence, the sub-LE
of the bidirectional triangle is then different from the sub-LEs of
the networks with input from one laser. As soon as the bidirectional
triangle synchronizes, however, its sub-LE becomes equal to the
networks with input from one laser. The reason is that the signals
from the two other laser are superimposed coherently and thus equal
the signal from one laser.

Fig.~\ref{F_V_E_1}(b) shows the two distinct sub-LEs of the inner and
outer lasers of a chain of three bidirectionally coupled lasers in
comparison with the sub-LE of the bidirectional pair. If the coupling
strength $\sigma$ is large enough for the outer lasers to synchronize,
as seen in Fig.~\ref{F_V_E_1}(c), then the inner laser receives the
coherent superposition of the signals of the outer lasers. Thus, all
three lasers in the chain effectively receive input from one
unsynchronized laser. In consequence, the chain can be reduced to a
bidirectional pair of unsynchronized lasers. Indeed,
Fig.~\ref{F_V_E_1}(b) confirms that the sub-LEs of the outer and inner
lasers of the chain become identical and also equal to the sub-LE of
the bidirectional pair in this regime.

\begin{figure}
  \begin{minipage}{0.44\columnwidth}
    \includegraphics[width=\columnwidth]{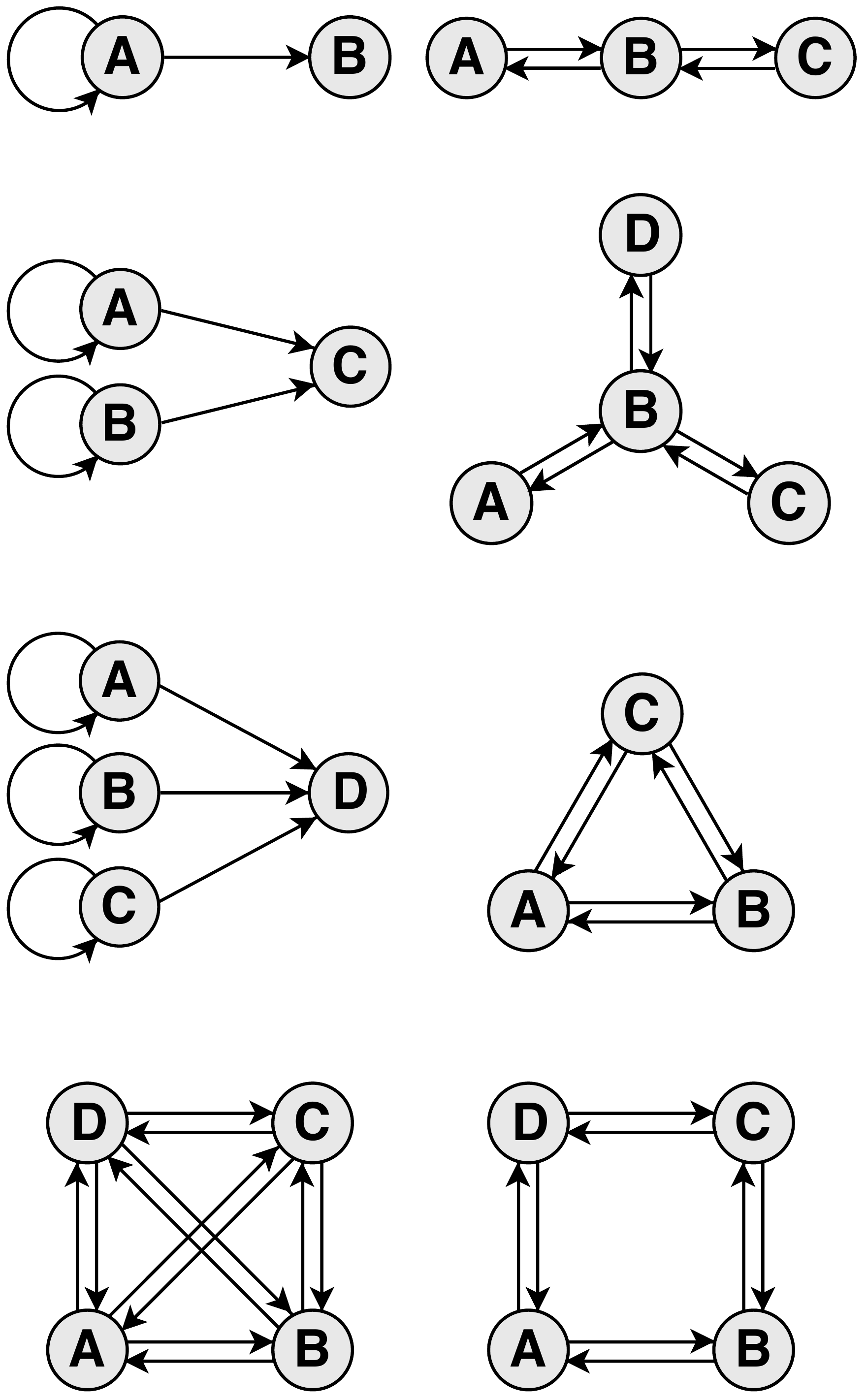}
  \end{minipage}
  \begin{minipage}{0.54\columnwidth}
    \includegraphics[width=\columnwidth]{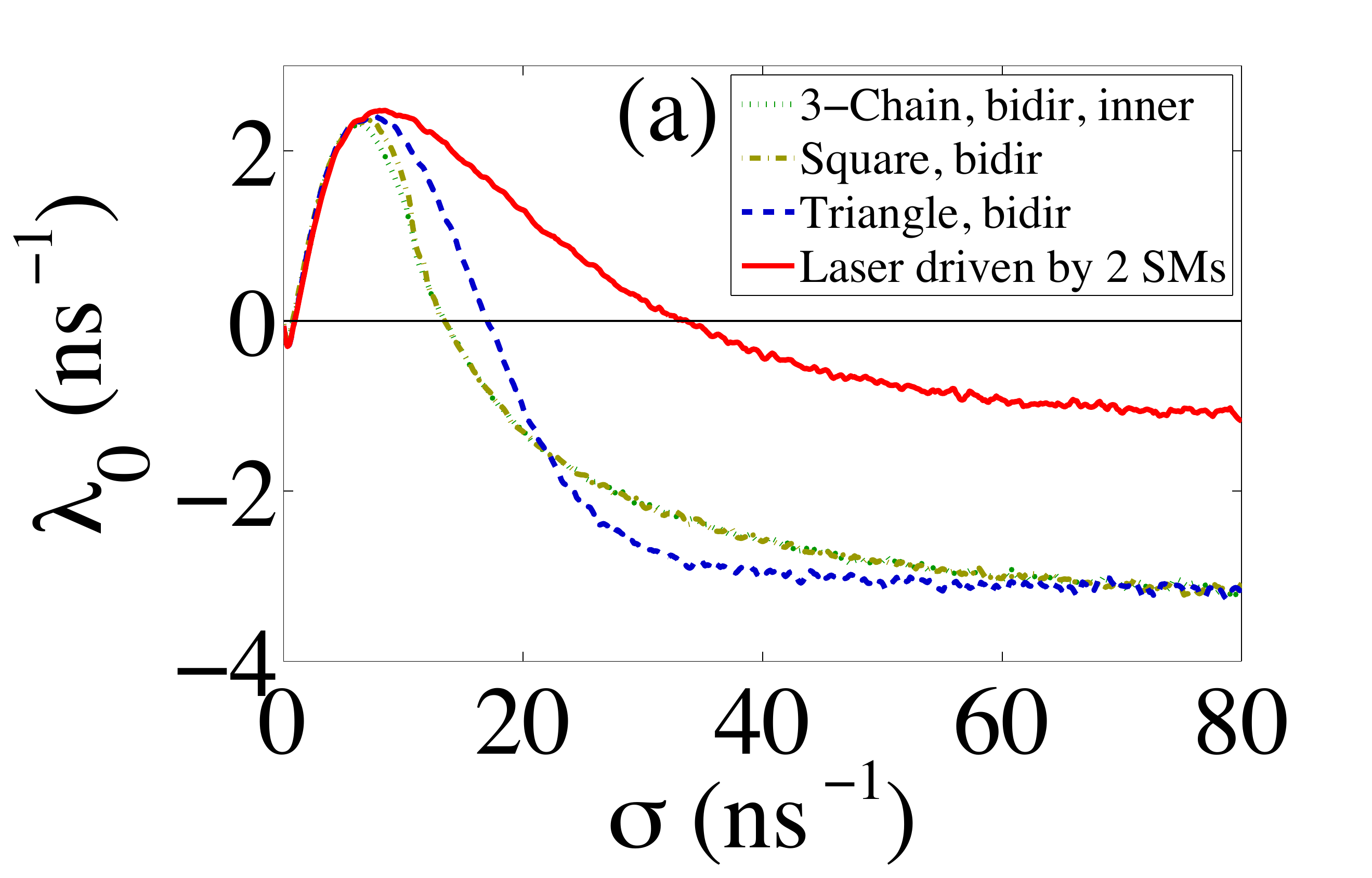}
    \newline
    \includegraphics[width=\columnwidth]{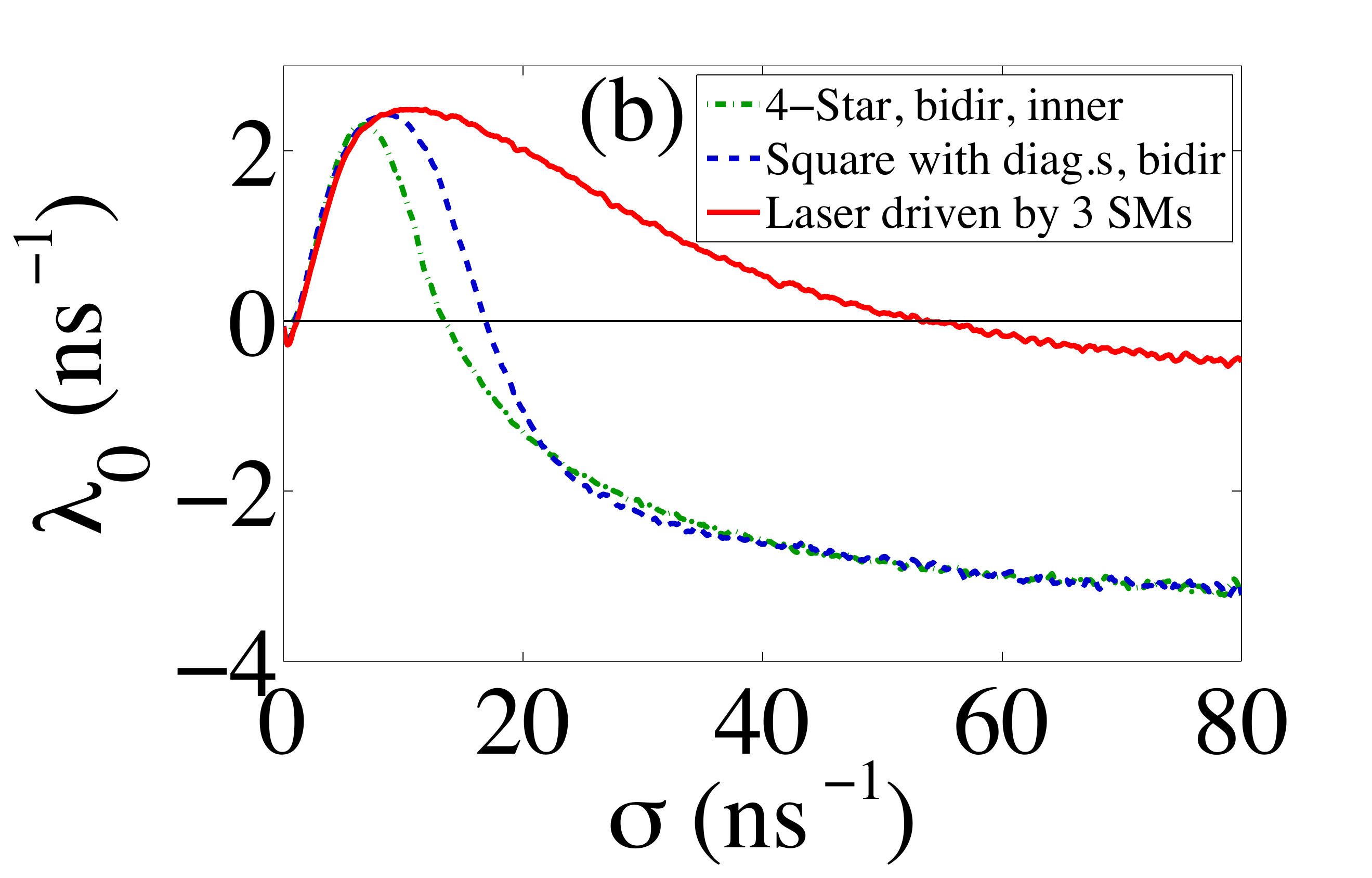}
    \newline
    \includegraphics[width=\columnwidth]{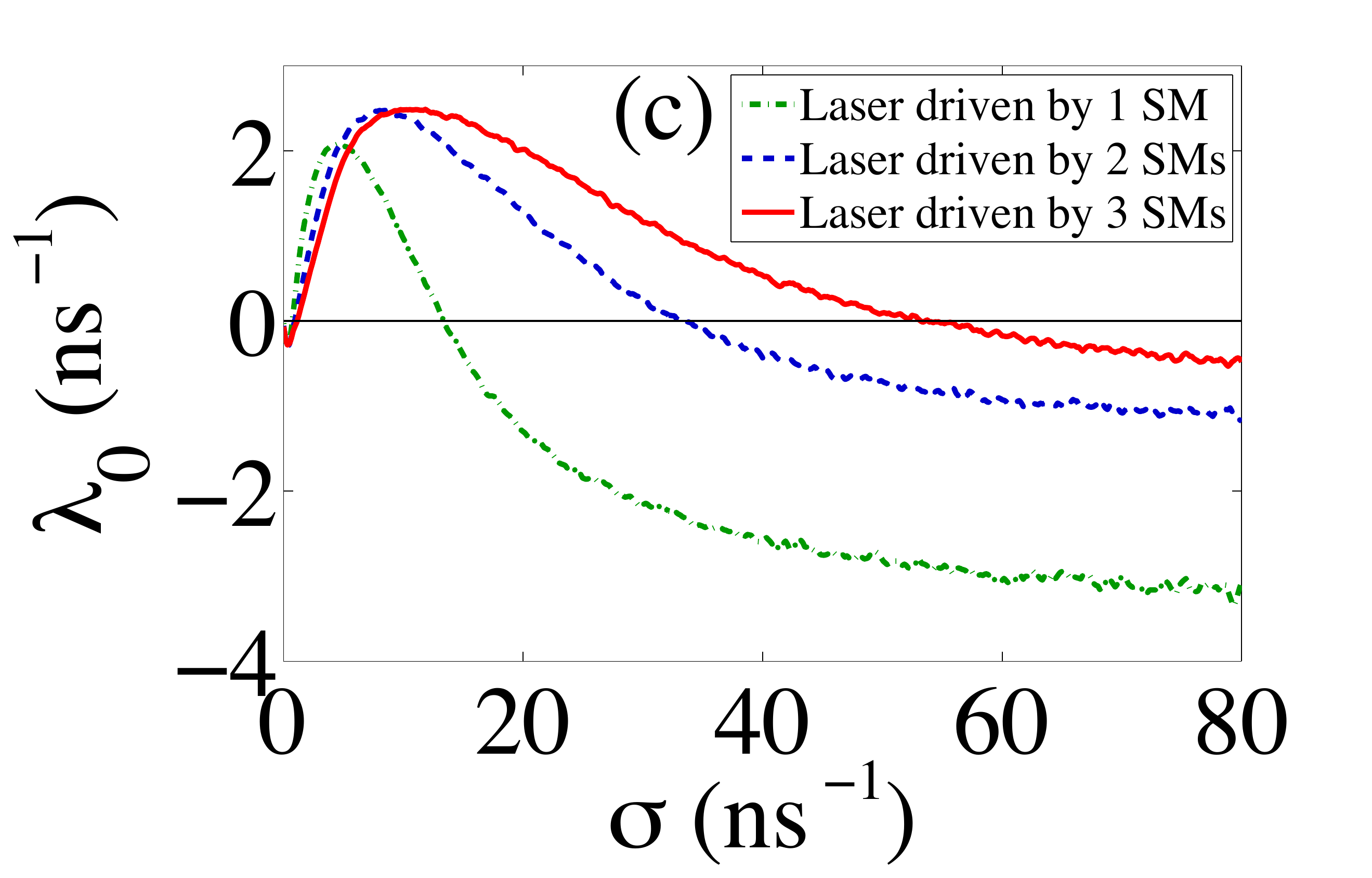}
  \end{minipage}
  \caption{(Color online) (a)~Sub-LEs $\lambda_{0}$ for several
    network patterns where the lasers receive input from two lasers
    (see left) for a coupling delay time $\tau=10\,\mathrm{ns}$ vs
    coupling strength $\sigma$. (b)~Sub-LEs $\lambda_{0}$ for several
    network patterns where the lasers receive input from three lasers
    (see left) for a coupling delay time $\tau=10\,\mathrm{ns}$ vs
    coupling strength $\sigma$. (c)~Sub-LEs $\lambda_{0}$ of a laser
    driven by one, two or three independent laser with self-feedback
    for a coupling delay time $\tau=10\,\mathrm{ns}$ vs coupling
    strength $\sigma$.}
  \label{F_V_E_2}
\end{figure}

Fig.~\ref{F_V_E_2}(a) shows a comparison of the sub-LEs of several
network configurations (depicted on the left side) in which the lasers
receive input from two other lasers. Fig.~\ref{F_V_E_2}(b) shows a
comparison of the sub-LEs of several networks (depicted on the left
side) where the lasers receive input from three other lasers. In both
diagrams it turns out that for small coupling strengths in a partial
range of the strong chaos regime, the sub-LE indeed only depends on
the number of inputs and is equal for networks which have the same
number of inputs. For larger coupling strengths, however, the sub-LEs
differ from each other. The reason for this is that the two or three
inputs correspond to an incoherent superposition of two or three laser
signals. The level of coherence or incoherence depends on the topology
of the subnetwork from which the signals originate. These different
levels of incoherence play a larger role for the laser receiving the
signals for increasing coupling strength.

Fig.~\ref{F_V_E_2}(c) shows a comparison of the sub-LEs of a laser
which receives unidirectional input from one, two or three mutually
uncoupled lasers with self-feedback. The influence of the topology of
the subnetwork from which the laser receives the two or three signals
is eliminated by the fact that the two or three driving lasers are
completely independent from each other. Also, the driven laser does
not feed back any signal to the driving lasers. Hence,
Fig.~\ref{F_V_E_2}(c) shows the dependence of the sub-LE on the number
of the completely incoherently superimposed inputs. For increasing
number of inputs, the statistics of the received summed signal gets
less similar to the one of a laser and increasingly similar to
noise. We observe that with increasing number of inputs, the sub-LE
gets larger for large coupling strengths $\sigma$ and smaller for
small $\sigma$.

\section{Summary}

In this paper we have extended the investigations on strong and weak
chaos of \cite{Heiligenthal2011} by focusing on the dynamics of
semiconductor lasers. After describing the LK equations and their
linearization as a model for the numerical simulation of a
semiconductor laser with time-delayed couplings, we have introduced
the artificial sub-LE $\lambda_0$ and have explained how to determine
its sign by experiments.

Strong and weak chaos show different scaling properties of the maximal
LE with the delay time. The sign of the sub-LE $\lambda_0$
distinguishes between strong and weak chaos. The transition sequence
`weak to strong chaos and back to weak chaos' upon monotonically
increasing the coupling strength $\sigma$ of a single laser's
self-feedback has been shown for numerical calculations of the LK
equations. At the transition between strong and weak chaos, the sub-LE
vanishes, $\lambda_0=0$. At this transition we found
$\lambda_{\mathrm{m}}\,\tau\sim\sqrt{\tau}$. Transitions between
strong and weak chaos by changing $\sigma$ could also be found for the
Rössler and Lorenz dynamics.

Counterintuitively, the difference between strong and weak chaos is
not directly visible from the trajectory although the difference of
the trajectories induces the transitions between the two types of
chaos. In addition, a linear measure like the auto-correlation
function cannot unambiguously reveal the difference between strong and
weak chaos, either. Although the auto-correlations after one delay
time are significantly higher for weak chaos than for strong chaos, it
was not possible to detect a qualitative difference. But we could
relate the trajectories of strong and weak chaos to the properties of
the external cavity modes of the laser. If two time-scale separated
self-feedbacks are present, the shorter feedback has to be taken into
account for the definition of a new sub-LE $\lambda_{0,s}$ which in
this case determines the occurence of strong or weak chaos. We have
shown that the sub-LE scales with the square root of the effective
pump current $\sqrt{p-1}$, as well in its magnitude as in the position
of the critical coupling strengths.

For networks of delay-coupled lasers, we explained using the master
stability formalism the condition
$|\gamma_k|<\e^{-\lambda_{\mathrm{m}}\,\tau}$ for stable chaos
synchronization. Hence, synchronization of any network depends only on
the properties of a single laser and the eigenvalue gap of the
coupling matrix. We refined the master stability function for the LK
dynamics to allow for precise practical prediction of
synchronization. We provided the first experimental evidence of strong
and weak chaos in bidirectionally delay-coupled lasers which supports
the sequence `weak to strong to weak chaos'. For networks with several
distinct sub-LEs it has been shown that the maximal sub-LE of the
network determines whether the network's maximal LE scales strongly or
weakly with increasing delay time. As a consequence, complete
synchronization of a network is excluded for arbitrary networks which
contain at least one strongly chaotic laser. Finally, we showed that
the sub-LE of a driven laser depends on the number of the incoherently
superimposed inputs from desynchronized input lasers.

\section*{Acknowledgments}

We thank the Deutsche Forschungsgemeinschaft and the
Leibniz-Rechenzentrum in Garching, Germany, for their support of this
work. M.\,C.\,S. and I.\,F. acknowledge the support by MICINN (Spain)
under project \mbox{DeCoDicA} (TEC2009-14101).

\appendix

\section{Parameters for the simulation of the Lang-Kobayashi
  equations}

If not mentioned differently in the text, the constants listed in
Tab.~\ref{T1} were used in the simulation of the LK equations.

\begin{table}[h]
  \caption{Used constants in the simulation of the LK equations.
    Values are taken from \cite{Ahlers1998}\label{T1}.}

  \begin{tabular}{lcl}
    \midrule
    \midrule 
    Parameter & Symbol & Value\\
    \midrule
    Linewidth enhancement factor & $\alpha$ & 5\\
    Differential optical gain & $G_{\mathrm{N}}$ & $2.142\times10^{4}\,\mathrm{s}^{-1}$\\
    Laser frequency & $\omega_{0}$ & $2\pi\, c/(635\,\mathrm{nm})$\\
    Pump current relative to $J_{\mathrm{th}}$ & $p$ & $1.02$\\
    Threshold pump current & &\\
    of solitary laser & $J_{\mathrm{th}}$ & $\gamma\, N_{\mathrm{sol}}$\\
    Carrier decay rate & $\gamma$ & $0.909\times10^{9}\,\mathrm{s}^{-1}$\\
    Carrier number of solitary laser & $N_{\mathrm{sol}}$ & $1.707\times10^{8}$\\
    Cavity decay rate & $\Gamma$ & $0.357\times10^{12}\,\mathrm{s}^{-1}$\\
    \midrule
    \midrule
  \end{tabular}
\end{table}

\vspace{1cm}

\section{Critical coupling strengths in dependence on the pump
  current}

Tab.~\ref{T2} lists the critical coupling strengths for which
$\lambda_0 = 0$ and at which the transitions from weak to strong chaos
and from strong to weak chaos appear, in dependence on the pump
current $p$.

\begin{table}[h]
  \caption{Critical coupling strengths where $\lambda_0 = 0$ and at which
    the transitions from weak to strong chaos ($\sigma_{\mathrm{crit,1}}$)
    and from strong to weak chaos ($\sigma_{\mathrm{crit,2}}$) happen in
    dependence on the pump current $p$\label{T2}.}

  \begin{tabular}{ccc}
    \midrule
    \midrule 
    $p$ & $\sigma_{\mathrm{crit,1}}\,(\mathrm{ns^{-1}})$ & $\sigma_{\mathrm{crit,2}}\,(\mathrm{ns^{-1}})$\\
    \midrule
    1.02 & 0.80 & 13.44\\
    1.05 & 0.96 & 20.80\\
    1.10 & 1.44 & 29.28\\
    1.15 & 1.76 & 35.68\\
    1.20 & 2.08 & 40.64\\
    1.25 & 2.40 & 45.12\\
    1.30 & 2.72 & 48.96\\
    1.35 & 2.88 & 52.64\\
    1.40 & 3.20 & 56.16\\
    1.45 & 3.36 & 58.88\\
    1.50 & 3.68 & 61.92\\
    \midrule
    \midrule
  \end{tabular}
\end{table}


%

\end{document}